\documentclass[10pt, letterpaper, onecolumn]{article}

\usepackage[top=1in, bottom=1in, left=1.5in, right=1.5in]{geometry}

\usepackage{mathptmx}
\usepackage[T1]{fontenc}

\usepackage{setspace}
\usepackage{xcolor}
\usepackage{amsmath,amssymb}
\usepackage{tikz}
\usetikzlibrary{positioning,arrows.meta,calc}
\usepackage{enumitem}
\usepackage[hidelinks, colorlinks=true, urlcolor=blue, citecolor=blue, linkcolor=blue]{hyperref}
\usepackage{natbib}
\usepackage{floatpag}

\usepackage{graphicx}
\usepackage{xspace}

\definecolor{stepfill}{HTML}{ECECEC}
\definecolor{stepline}{HTML}{7A7A7A}
\definecolor{decfill}{HTML}{E8E3F5}
\definecolor{decline}{HTML}{6E56A6}
\definecolor{failfill}{HTML}{F4E1E7}
\definecolor{failline}{HTML}{C96987}
\definecolor{testfill}{HTML}{DCF0EE}
\definecolor{testline}{HTML}{2E8B84}

\tikzset{
  box/.style={draw, rounded corners=3pt, align=center, minimum height=1.1cm,
              inner sep=4pt, line width=0.6pt},
  lbox/.style={box, text width=6.0cm},
  rbox/.style={box, text width=5.2cm},
  stp/.style={fill=stepfill, draw=stepline},
  dcn/.style={fill=decfill, draw=decline},
  flt/.style={fill=failfill, draw=failline},
  tst/.style={fill=testfill, draw=testline},
  arr/.style={-{Stealth[length=2mm]}, line width=0.6pt, draw=black!60},
  lab/.style={font=\scriptsize, fill=white, inner sep=1pt},
  key/.style={draw, rounded corners=2pt, minimum width=0.35cm, minimum height=0.35cm,
              inner sep=0pt, line width=0.5pt},
}
\newcommand{\bt}[2]{{\footnotesize\bfseries #1}\\[1pt]{\scriptsize #2}}

\newcommand{\Sec}[1]{{\protect\hyperref[sec:#1]{Section~\ref*{sec:#1}}}}

\newcommand{\Fig}[1]{{\protect\hyperref[fig:#1]{Figure~\ref*{fig:#1}}}}

\newcommand{\subFig}[2]{{\protect\hyperref[fig:#1]{Figure~\ref*{fig:#1}~#2}}}
\newcommand{\Equ}[1]{{\protect\hyperref[equ:#1]{Equation~\ref*{equ:#1}}}}

\newcommand{\Tab}[1]{{\protect\hyperref[tab:#1]{Table~\ref*{tab:#1}}}}

\newcommand{\App}[1]{{\protect\hyperref[app:#1]{Appendix~\ref*{app:#1}}}}

\newcommand{\PTED}{\textsc{pted}\xspace}

\title{PTED: A multi-dimensional two-sample test for scientific inference and generative machine learning}
\author{Connor Stone\footnote{\href{mailto:connor.stone@utoronto.ca}{connor.stone@utoronto.ca}} \\ \textit{University of Toronto}}
\date{} 

\begin{document}

\maketitle

\begin{abstract}
Two-sample tests are widely applicable in inference and generative modelling, yet users frequently fall back on heuristics and visual inspection due to lack of an accessible test that operates in multiple dimensions.
I present Permutation Test using the Energy Distance (\PTED), a Python implementation of a powerful two-sample test described in \citet{szekely2004testing}. 
The test statistic is the energy distance, a metric on probability distributions that admits a natural sample estimator built from pairwise distances.
A permutation test is run on energy distances to produce an exact two-sample test.
Because the statistic depends on the data only through pairwise distances, the test applies in high dimensions, on learned feature representations, at large or small or imbalanced sample sizes, and to any data type on which a distance can be defined.
An approximation to the formula allows \PTED to scale linearly with both number of dimensions and samples while retaining most discriminative power.
I highlight a number of instructive sensitivity tests on \PTED alongside other multi-dimensional two-sample tests for comparison\footnote{All tests in this manuscript may be reproduced using the scripts here: \href{https://github.com/ConnorStoneAstro/pted\_tests}{https://github.com/ConnorStoneAstro/pted\_tests}}.
\PTED is the only method that is sensitive to all tests, though conventional Maximum Mean Discrepancy is identical for all but an over-fitting test.
\end{abstract}

\section{Introduction} \label{sec:intro} 

A two-sample test compares two collections of data points in order to assess whether they were drawn from the same probability distribution.
Such a simple problem statement appears broadly in scientific inference and generative machine learning.
Most two-sample tests are framed as null hypothesis significance tests: one computes a $p$-value under the null hypothesis $H_0$ that both samples were drawn from the same distribution~\citep{moore2009introduction}.
For a test statistic $T$ with observed value $t$, the $p$-value is $P_{H_0}(T\geq t)$: the probability, under the null, of a statistic at least as extreme as the one observed.
The most widely known two-sample test algorithm is the Kolmogorov-Smirnov test (KS-test) which uses the maximum discrepancy in a cumulative distribution function (CDF) under the null to compute a $p$-value~\citep{an1933sulla, simard2011computing}.

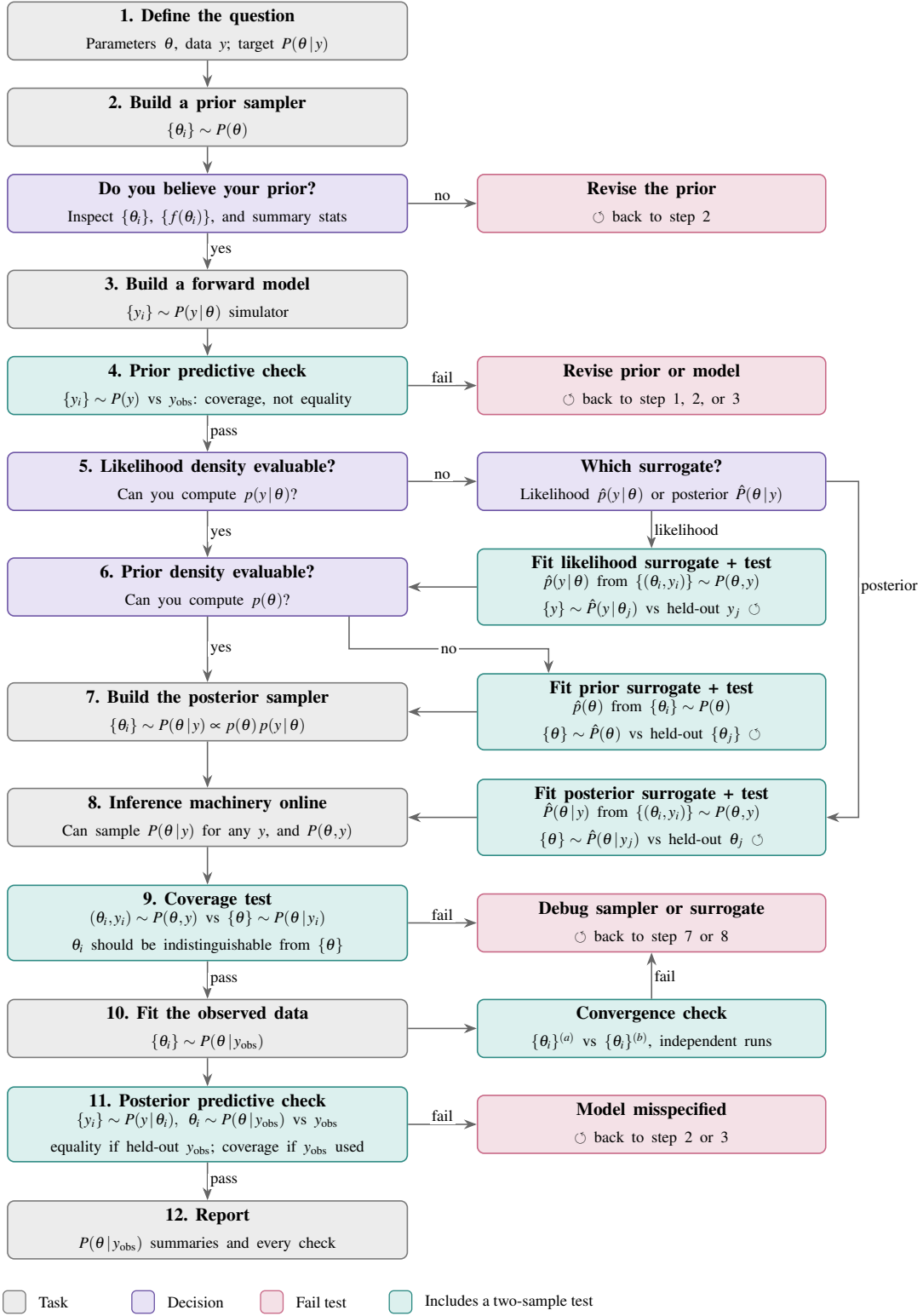
\begin{figure}[p]
\thisfloatpagestyle{empty}
\begin{center}
\thickmuskip=5mu \medmuskip=4mu
\renewcommand{\bt}[2]{{\footnotesize\bfseries #1}\\[0pt]{\scriptsize #2}}
\newcounter{nodetipcnt}
\ifdefined\pdfannot \ifdefined\pdfstringdef
  \newcommand{\nodetip}[2]{%
    \stepcounter{nodetipcnt}%
    \pdfstringdef\nodetipstr{#2}%
    \path let \p1=($(#1.north east)-(#1.south west)$) in
      node[inner sep=0pt, outer sep=0pt] at (#1)
      {\pdfannot width \x1 height \y1 depth 0pt
         {/Subtype /Widget /FT /Btn /Ff 65536 /F 768 /H /N
          /BS << /W 0 >> /C [ ]
          /T (nodetip\thenodetipcnt) /TU (\nodetipstr)}%
       \rule{\x1}{0pt}\rule{0pt}{\y1}};}
\fi\fi
\ifdefined\nodetip\else \newcommand{\nodetip}[2]{}\fi
\begin{tikzpicture}[x=1cm,y=1cm, box/.append style={minimum height=0.9cm, inner sep=3pt}]
\node[lbox,stp] (s1)  at (0,  0.0) {\bt{1. Define the question}{Parameters $\theta$, data $y$; target $P(\theta\,|\,y)$}};
\node[lbox,stp] (s2)  at (0, -1.34) {\bt{2. Build a prior sampler}{$\{\theta_i\} \sim P(\theta)$}};
\node[lbox,dcn] (q1)  at (0, -2.68) {\bt{Do you believe your prior?}{Inspect $\{\theta_i\}$, $\{f(\theta_i)\}$, and summary stats}};
\node[lbox,stp] (s3)  at (0, -4.17) {\bt{3. Build a forward model}{$\{y_i\} \sim P(y\,|\,\theta)$ simulator}};
\node[lbox,tst] (s4)  at (0, -5.51) {\bt{4. Prior predictive check}{$\{y_i\} \sim P(y)$ vs $y_{\rm obs}$: coverage, not equality}};
\node[lbox,dcn] (q5)  at (0, -7.0) {\bt{5. Likelihood density evaluable?}{Can you compute $p(y\,|\,\theta)$?}};
\node[lbox,dcn] (q6)  at (0, -8.64) {\bt{6. Prior density evaluable?}{Can you compute $p(\theta)$?}};
\node[lbox,stp] (s7)  at (0,-10.58) {\bt{7. Build the posterior sampler}{$\{\theta_i\} \sim P(\theta\,|\,y) \propto p(\theta)\,p(y\,|\,\theta)$}};
\node[lbox,stp] (s8)  at (0,-12.22) {\bt{8. Inference machinery online}{Can sample $P(\theta\,|\,y)$ for any $y$, and $P(\theta, y)$}};
\node[lbox,tst] (s9)  at (0,-13.86) {\bt{9. Coverage test}{$(\theta_i, y_i) \sim P(\theta, y)$ vs $\{\theta\} \sim P(\theta\,|\,y_i)$\\ $\theta_i$ should be indistinguishable from $\{\theta\}$}};
\node[lbox,stp] (s10) at (0,-15.5) {\bt{10. Fit the observed data}{$\{\theta_i\} \sim P(\theta\,|\,y_{\rm obs})$}};
\node[lbox,tst] (s11) at (0,-16.99) {\bt{11. Posterior predictive check}{$\{y_i\} \sim P(y\,|\,\theta_i),\ \theta_i \sim P(\theta\,|\,y_{\rm obs})$ vs $y_{\rm obs}$\\ equality if held-out $y_{\rm obs}$; coverage if $y_{\rm obs}$ used}};
\node[lbox,stp] (s12) at (0,-18.63) {\bt{12. Report}{$P(\theta\,|\,y_{\rm obs})$ summaries and every check}};
 
\node[rbox,flt] (f1)  at (6.9, -2.68) {\bt{Revise the prior}{$\circlearrowleft$ back to step 2}};
\node[rbox,flt] (f4)  at (6.9, -5.51) {\bt{Revise prior or model}{$\circlearrowleft$ back to step 1, 2, or 3}};
\node[rbox,dcn] (qs)  at (6.9, -7.0) {\bt{Which surrogate?}{Likelihood $\hat p(y\,|\,\theta)$ or posterior $\hat P(\theta\,|\,y)$}};
\node[rbox,tst] (ls)  at (6.9, -8.64) {\bt{Fit likelihood surrogate + test}{$\hat p(y\,|\,\theta)$ from $\{(\theta_i, y_i)\} \sim P(\theta, y)$\\ $\{y\} \sim \hat P(y\,|\,\theta_j)$ vs held-out $y_j$ $\circlearrowleft$}};
\node[rbox,tst] (ps)  at (6.9,-10.58) {\bt{Fit prior surrogate + test}{$\hat p(\theta)$ from $\{\theta_i\} \sim P(\theta)$\\ $\{\theta\} \sim \hat P(\theta)$ vs held-out $\{\theta_j\}$ $\circlearrowleft$}};
\node[rbox,tst] (sp)  at (6.9,-12.22) {\bt{Fit posterior surrogate + test}{$\hat P(\theta\,|\,y)$ from $\{(\theta_i, y_i)\} \sim P(\theta, y)$\\ $\{\theta\} \sim \hat P(\theta\,|\,y_j)$ vs held-out $\theta_j$ $\circlearrowleft$}};
\node[rbox,flt] (f9)  at (6.9,-13.86) {\bt{Debug sampler or surrogate}{$\circlearrowleft$ back to step 7 or 8}};
\node[rbox,tst] (cv)  at (6.9,-15.5) {\bt{Convergence check}{$\{\theta_i\}^{(a)}$ vs $\{\theta_i\}^{(b)}$, independent runs}};
\node[rbox,flt] (f11) at (6.9,-16.99) {\bt{Model misspecified}{$\circlearrowleft$ back to step 2 or 3}};
 
\draw[arr] (s1)  -- (s2);
\draw[arr] (s2)  -- (q1);
\draw[arr] (q1)  -- node[lab,right]{yes}  (s3);
\draw[arr] (s3)  -- (s4);
\draw[arr] (s4)  -- node[lab,right]{pass} (q5);
\draw[arr] (q5)  -- node[lab,right]{yes}  (q6);
\draw[arr] (q6)  -- node[lab,right]{yes}  (s7);
\draw[arr] (s7)  -- (s8);
\draw[arr] (s8)  -- (s9);
\draw[arr] (s9)  -- node[lab,right]{pass} (s10);
\draw[arr] (s10) -- (s11);
\draw[arr] (s11) -- node[lab,right]{pass} (s12);
 
\draw[arr] (q1)  -- node[lab,above]{no}   (f1);
\draw[arr] (s4)  -- node[lab,above]{fail} (f4);
\draw[arr] (q5)  -- node[lab,above]{no}   (qs);
\draw[arr] (s9)  -- node[lab,above]{fail} (f9);
\draw[arr] (s10) -- (cv);
\draw[arr] (cv)  -- node[lab,right]{fail} (f9);
\draw[arr] (s11) -- node[lab,above]{fail} (f11);
 
\draw[arr] (qs) -- node[lab,right]{likelihood} (ls);
\draw[arr] (ls.west) -- (q6.east);
\draw[arr] ([xshift=2.2cm]q6.south) -- ++(0,-0.5) -| ([xshift=-1.6cm]ps.north);
\node[lab] at (3.75,-9.61) {no};
\draw[arr] (ps.west) -- (s7.east);
\path (qs.east) ++(0.5,0) coordinate (bpR);
\draw[arr] (qs.east) -- (bpR) |- (sp.east);
\node[lab,right] at (bpR |- ls) {posterior};
\draw[arr] (sp.west) -- (s8.east);
 
\nodetip{s1}{Try to answer the question "what do I want to know?" and also "what data do I have?" then build a Directed Acyclic Graph (DAG) between them that includes any intermediate model parameters that need to be added to make that link. Your theta is then all the values in the DAG except the data values, which are y. Every later step depends on the connection between theta and y being explicit. Note that you don't need to simulate the whole universe to answer every question, but you do need a prior over everything in theta so it needs to be filled with variables you understand. Also, you may wish to work with summary statistics S(y) instead of the full data y, which is ok, but for the sake of the DAG, its best to go right to the data y.}
\nodetip{s2}{Write a function that returns random draws of theta. This is always possible: uniform over physical bounds, a previous experiment's posterior, or resampling from a catalogue or simulation bank. You do not need to evaluate the density yet.}
\nodetip{q1}{Draw many samples. Plot marginals, parameter pairs, derived physical quantities, and summary statistics. Would you bet against regions the prior visits often? Does it never visit regions you find plausible? If you can dismiss a sample out-of-hand without even seeing data, then the prior sampler is wrong. Fix it now: a bad prior pulls the posterior whenever the data are weak. Be extra wary in high dimensions where volume effects can dominate.}
\nodetip{s3}{Write a simulator from theta to y in the format of the real data: deterministic physics plus every stochastic step (noise, background, detector response, selection). A simulator can be built even when a closed-form likelihood cannot. Your DAG from step 1 will be helpful here.}
\nodetip{s4}{Draw theta from the prior and then draw y from the forward model given theta, doing this many times produces prior predictive samples. This is also how you draw joint samples P(theta, y). The prior predictive should contain the real data, not necessarily match it, because it mixes over the whole prior. If it appears that the prior predictive will never sample anything data-like, then something is wrong and you need to revise the model. Use an asymmetric two-sample test here rather than checking for equality, you check for containment.}
\nodetip{q5}{Can you write a function that returns the (log) likelihood density up to a constant? Does it run fast enough for later sampling? Are you able to compute gradients of the likelihood? If a reliable likelihood density is available and fast it should almost always be preferred over a fitted surrogate. If the simulator has internal random steps you cannot integrate out analytically, or is trans-dimensional, or is exceedingly slow to compute, it may not be viable for inference.}
\nodetip{q6}{Can you write a function that returns the prior density? Does it run fast enough for later sampling? Are you able to compute gradients of the prior? A fast and density based prior should be preferred over a fitted surrogate, but only if it fully encodes prior knowledge and does not allow impossible/unbelievable parts of parameter space.}
\nodetip{s7}{Combine the evaluable prior and likelihood densities into a sampler that returns posterior draws of theta for any input data y. The densities may be known only up to a normalizing constant. In most cases, some variant of a Markov-Chain Monte-Carlo will be needed to build the sampler.}
\nodetip{s8}{Merge point. Whichever path you took you now have a black box: y in, posterior samples of theta out. With the prior sampler and forward model from steps 2 and 3 you can also draw from the joint distribution of theta and y. Everything after this tests the box's behaviour, not its internals.}
\nodetip{s9}{Repeat many times: draw theta from the prior, forward model y from it, run the posterior sampler on that y, and record where the true theta falls among the posterior samples. The ranks must be uniform. A failure is an inference bug, not a physics problem, because the true values and data were generated from the same model being tested.}
\nodetip{s10}{Run the posterior sampler on the real data. Run more than one chain from independent seeds or starting points so the convergence check has at least two sample sets to compare. Multiple runs of the posterior sampler must produce statistically indistinguishable results or the sampler is not converged.}
\nodetip{s11}{For each posterior draw of theta, use the forward model to simulate a replicated dataset y and compare with the observed data. Ideally, compare the posterior predictive data with held-out real y data, then it is possible to run a true two-sample test. If you cannot use held-out data, then it is still possible to check for containment. If the posterior predictive cannot produce results that statistically match the real data, then a failure means the model, not the code, is wrong.}
\nodetip{s12}{Report marginals (corner plot), credible intervals, and correlations of the posterior, possibly by simply reporting the collection of posterior samples. For extra reporting strength include the result for every check passed: prior predictive, surrogate tests, coverage, convergence, posterior predictive, and sensitivity of the result to the prior.}

\nodetip{f1}{Return to step 2 and change the prior: widen or narrow ranges, add a correlation, or use a better source of information. Re-check before continuing. If you don't believe your prior then it is not your prior.}
\nodetip{f4}{No prior draw produces data like the real data. Either the prior excludes the truth (step 2), the forward model is missing physics or noise (step 3), or the question itself was framed incorrectly (step 1).}
\nodetip{qs}{Two mutually exclusive options. A likelihood surrogate gives an evaluable likelihood density, this is beneficial as the prior may be adjusted later and it is also typically easier to fit a likelihood function (often reduces to a Gaussian mixture). May be fit using any distribution on theta (only needs the forward model). A posterior surrogate replaces the whole density path, this is beneficial as it can be very fast to sample and run many tests on, however it is often harder to fit and the prior is then locked in. Must be fit exclusively on joint samples. There is no reason to fit both as the posterior surrogate subsumes the likelihood.}
\nodetip{ls}{Fit a flexible model of the likelihood density to forward model simulated theta and y pairs (where theta may come from any distribution, but the prior is a good start). Use hold out simulations, (theta, y) pairs that weren't used for fitting, to compare with y draws from the surrogate at given theta, use a two-sample test to compare. Repeat fitting until it passes, this indicates that the model has encoded the forward model up to the resolution of the two-sample test. Can also run a two-tailed two-sample test between generated samples and the training samples to check for over-fitting.}
\nodetip{ps}{Fit a flexible density model to the prior sample bank. Hold out part of the bank, draw fresh theta from the surrogate, and two-sample test against the held-out samples. Also run a two-tailed test on prior surrogate samples vs the training samples: being too close to the training data means replication/memorization. Only needed if the training sample is finite and had to be re-used over many epochs.}
\nodetip{sp}{Fit a flexible conditional model that returns theta samples (and ideally a density) given y, trained on simulated joint theta and y pairs. Validate with a two-sample test on held-out pairs: surrogate draws at a held-out y against the theta that generated it. Also run a two-tailed two-sample test on surrogate posterior samples vs the training samples, being too close to the training data means replication/memorization. Only needed if the training sample is finite and had to be re-used over many epochs.}
\nodetip{f9}{The posterior box is miscalibrated: overconfident, under-confident or biased. The test data came from the model, so fix the sampler (step 7) or refit the surrogate (step 8). No physics or reality-matching is tested here since the truth comes from the same model being tested, this test is only concerned that the inference machinery is being run to convergence (and not getting hung up on a bug, incomplete integral, unconverged chain, etc.).}
\nodetip{cv}{Two or more independent runs of the posterior sampler on the real data should give indistinguishable theta samples. If they differ, at least one run has not properly explored the posterior. If step 9 passed, this may mean that the data is out of distribution and the sampler is struggling. For a start, recheck the inference machinery with a more comprehensive coverage test (step 9).}
\nodetip{f11}{Coverage and convergence tests have passed at this point, so the inference code is solid leaving only that the model is wrong. Some physics is missing from the forward model (step 3) or the prior excludes the truth (step 2).}
 
\node[key,stp] (k1) at (-3.0,-19.75) {};
\node[right=2pt of k1,font=\scriptsize] {Task};
\node[key,dcn] (k2) at (-1.0,-19.75) {};
\node[right=2pt of k2,font=\scriptsize] {Decision};
\node[key,flt] (k3) at (1.0,-19.75) {};
\node[right=2pt of k3,font=\scriptsize] {Fail test};
\node[key,tst] (k4) at (3.0,-19.75) {};
\node[right=2pt of k4,font=\scriptsize] {Includes a two-sample test};
\end{tikzpicture}
\end{center}
    \caption{Inference flowchart including locations of two-sample tests as green blocks. Notation: $\{x_i\} \sim P(\cdot)$ draws from a sampler; $p(\cdot)$ an evaluable density; $\hat{p}, \hat{P}$ a fitted surrogate;\\ $y_{\rm obs}$ the real data; $j$ indexes held-out samples not used for fitting.}
    \label{fig:inference}
\end{figure}

Two-sample tests appear frequently in scientific inference and are central to modern generative modelling.
\Fig{inference} presents a modern flowchart for robust inference including positions where generative/surrogate models may be applied.
Seven of its steps suggest a two-sample test as a validation check before the analysis proceeds.
For example, the classic coverage test in step 9, which involves first drawing joint samples $(\theta_i,y_i)$ from the prior/forward model, then testing if each $\theta_i$ is distinguishable from samples obtained via the posterior machinery $P(\theta\,|\,y_i)$.
Three of the possible two-sample tests in the flowchart are due to the inclusion of generative models.
The ultimate goal of a generative machine learning model is explicitly to produce new data drawn from the same probability distribution as its training data.
Simulation-based inference~\citep{Cranmer2020} makes this connection even clearer.
It works primarily with samples rather than densities and a greater effort is typically placed on validation.

The KS test is powerful and widely used, but is defined only for one-dimensional data; multivariate generalizations exist but have seen limited adoption~\citep{fasano1987multidimensional}.
While two-sample testing in multiple dimensions is commonly considered a challenge, in fact, the real challenge is choosing from a vast array of options.
Any projection of multiple dimensions into one dimension (such as projection along the first principal component axis) makes it possible to use the KS-test in many dimensions.
Further, any statistic that may be computed over two samples may be incorporated into a permutation test to produce an exact two-sample test.
For example, the Wasserstein distance~\citep{Panaretos2019}, Maximum Mean Discrepancy~\citep[MMD;][]{Gretton2006}, the Fr{\'e}chet Inception Distance~\citep[FID;][]{Heusel2017}, and the Feature Likelihood Divergence~\citep[FLD;][]{jiralerspong2023feature} all may be used in a permutation test.
Projection and permutation generate an infinite family of valid multivariate two-sample tests.

Here the \PTED software is presented, which implements one such two-sample test determined via a permutation test on the energy distance between samples of points.
This test has already been described in \citet{szekely2004testing} and is mathematically identical to MMD with a distance kernel $k(x,y) = -\lVert x - y\rVert$.
This contribution is a \texttt{Python} implementation, a technique for linear scaling with n-samples, and a number of instructive comparisons with other two-sample tests.
\PTED is highly competitive as a two-sample test in one dimension, and also in high-dimensional data relative to common tests used for generative models in machine learning.

\Sec{method} presents an overview of how \PTED works, what makes it so sensitive, and in what contexts.
\Sec{tests} includes comparison tests between \PTED and a number of competitive two-sample tests.
Finally, \Sec{conclusion} gives some discussion on the results of the tests and concludes with comments on best practices to use two-sample tests in common contexts.

\section{\PTED method}\label{sec:method}

\subsection{Core two-sample test}

The two-sample test problem may be stated more explicitly as follows.
Let $x = \{x_i\}_{i=1}^{n_x}$ be drawn i.i.d.\ from $F$ and $y = \{y_j\}_{j=1}^{n_y}$ i.i.d.\ from $G$, independently of one another, with $x_i, y_j \in \mathbb{R}^d$. 
A two-sample test assesses:
\begin{equation}
    H_0: F = G \qquad \text{against} \qquad H_1: F \neq G ,
    \label{equ:hypotheses}
\end{equation}
\noindent and \PTED does so using the energy distance as its test statistic.
The core \PTED algorithm is that of \citet{szekely2004testing}: a permutation test on the energy distance between two samples.
The energy distance between two distributions $F$ and $G$ is expressed as:

\begin{equation}
    D^2(F,G) = 2\,\mathbb{E}\!\left[\lVert X - Y \rVert\right]
             - \mathbb{E}\!\left[\lVert X - X' \rVert\right]
             - \mathbb{E}\!\left[\lVert Y - Y' \rVert\right],
    \label{equ:energydistance}
\end{equation}

\noindent where $D^2$ is the squared energy distance, $X, X' \sim F$ and $Y, Y' \sim G$ are independent copies, and $\lVert\cdot\rVert$ is the Euclidean norm ($\mathcal{L}_2$ norm) on $\mathbb{R}^d$.
A finite sample approximation of the energy distance may be determined naturally from pairwise distances between elements of each sample.
The corresponding sample statistic is obtained by replacing each expectation with an average over the available pairs,
\begin{equation}
    \hat{D}^2(x,y) = \frac{2}{n_x n_y}\sum_{i=1}^{n_x}\sum_{j=1}^{n_y} \lVert x_i - y_j \rVert
    - \frac{1}{n_x(n_x-1)}\sum_{i}\sum_{i' \neq i} \lVert x_i - x_{i'} \rVert
    - \frac{1}{n_y(n_y-1)}\sum_{j}\sum_{j' \neq j} \lVert y_j - y_{j'} \rVert ,
    \label{equ:samplestat}
\end{equation}
\noindent where the within-sample sums run over ordered pairs of distinct indices. 
Excluding the self-pairs $i = i'$ makes \Equ{samplestat} an unbiased estimator of $D^2(F,G)$, at the cost that it may take negative values on a finite sample; this is the convention adopted in \PTED (see \Sec{implementation}).

\citet{szekely2002statistics} notes that in one dimension the energy distance is twice the Cram{\'e}r distance.
The connection does not extend to higher dimensions; for example, the energy distance is rotation invariant\footnote{The energy distance is also translation invariant, which follows immediately from its definition in terms of differences.} while the Cram{\'e}r distance is not.
The energy distance is not invariant under a linear transformation of the parameter space of the samples.
If this is a desirable property then it is possible to standardize (whiten) the data, which is a common practice in machine learning.
A useful property of the energy distance is that it goes to zero if and only if $F=G$ making it a true metric in the space of distributions\footnote{$D(F,G)$ is also always non-negative, symmetric, and satisfies the triangle inequality. While we use $D^2(F,G)$ for our null hypothesis testing, the square root is a strictly monotonic function meaning our $p$-values are identical for $D(F,G)$ or $D^2(F,G)$.}.
Consequently, the \PTED test will always\footnote{When the sampling distributions have finite first moments, see \citet{szekely2004testing}.} distinguish two $F\neq G$ distributions as $n_x,n_y\to \infty$.
However, the number of samples needed to achieve a given sensitivity may be large.

\begin{figure*}
    \centering
    \includegraphics[width=\linewidth]{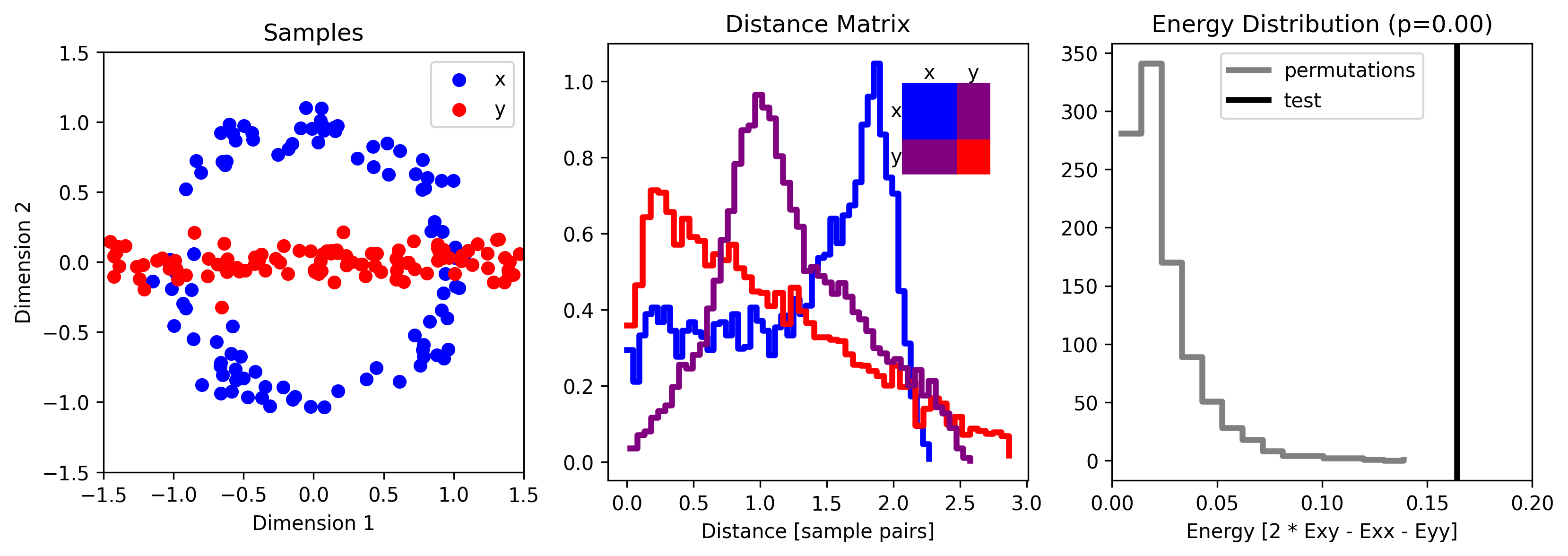}
    \includegraphics[width=\linewidth]{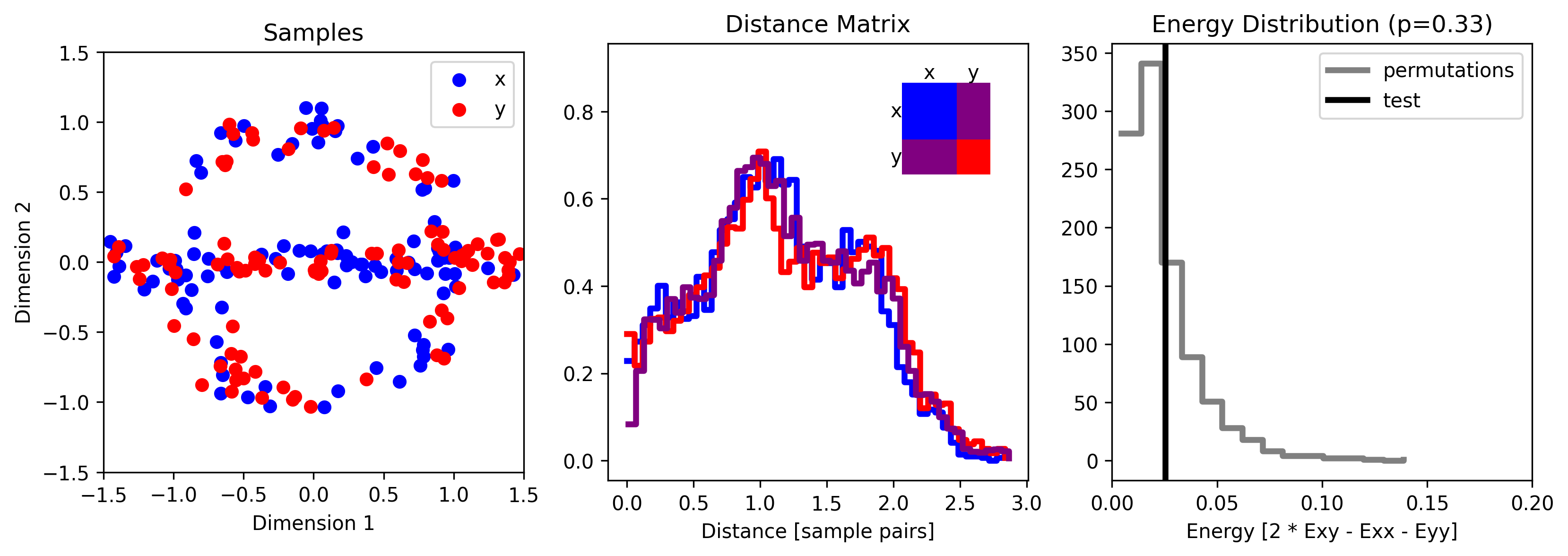}
    \caption{Visualization of the \PTED method. The blue points are the $x$ samples and the red points are the $y$ samples for the two-sample test. The left sub-figure plots the elements of the two samples. The centre sub-figure gives a histogram of the pairwise distances between elements from both samples, purple is used to represent the pairwise distances between $x$ and $y$ while the intra-sample pairwise distance histograms use the appropriate colours for each sample. A mock block histogram is inset which represents a distance matrix between the combined $x, y$ samples. The right sub-figure has a grey histogram of energy distances determined from many permutations (represented by the lower row), the black vertical line is the energy distance for the original labelling of $x,y$.}
    \label{fig:ptedmethod}
\end{figure*}

The energy distance between two finite samples is not directly interpretable as it lacks a reference scale.
However, it is possible to perform a permutation test on the energy distance between two samples, giving a highly meaningful $p$-value under the null hypothesis that $F=G$~\citep{good2013permutation, rizzo2019statistical}.
The permutation test involves repeatedly computing the energy distance under random re-labelling (permutation) of the elements between samples.
Let $z = (x_1,\dots,x_{n_x},y_1,\dots,y_{n_y})$ denote the pooled sample, $n = n_x+n_y$, and let $T(\pi) = \hat{D}^2$ evaluated on the first $n_x$ and last $n_y$ elements of $z$ after applying a permutation $\pi$ of its indices. 
Drawing $N$ permutations $\pi_1,\dots,\pi_N$ independently and uniformly from the symmetric group $S_n$, the test reports:
\begin{equation}
    \hat{p} = \frac{1 + \#\{k \le N : T(\pi_k) \ge T(\mathrm{id})\}}{N+1} .
    \label{equ:permp}
\end{equation}
\noindent Under $H_0$ the pooled sample is exchangeable, so $T(\mathrm{id})$ is distributed identically to each $T(\pi_k)$ and \Equ{permp} satisfies $\Pr_{H_0}(\hat{p} \le \alpha) \le \alpha$ for every $\alpha \in [0,1]$, with equality whenever $\alpha$ lies on the grid $\{1/(N+1), 2/(N+1), \dots, 1\}$. 
The test is therefore exact: its type-I error rate never exceeds the nominal level, for any $F$ and any sample size. 
Note that $\hat{p}$ is uniform on that grid rather than on the continuum, so the smallest attainable $p$-value is $1/(N+1)$.
A two-tailed version of the test is constructed as
\begin{equation}
    \hat{p}_{\rm two} = \min\!\left\{1,\; 2\min(\hat{p}_{\rm left}, \hat{p}_{\rm right})\right\},
    \label{equ:twotailed}
\end{equation}
\noindent where $\hat{p}_{\rm left}$ is computed as in \Equ{permp} except with $\#\{k \le N : T(\pi_k) \le T(\mathrm{id})\}$, and $\hat{p}_{\rm right}$ is computed exactly like \Equ{permp}.

\Fig{ptedmethod} gives a visual representation of how \PTED works.
In the top row it is visually clear the two samples are not drawn from the same distribution.
The columns represent the process of computing the energy distance for the $x,y$ samples.
The histograms in the centre column are not involved in the calculation of the energy distance, this visualization is for understanding only; instead \PTED takes an expectation over the distances. 
Since the two samples are different, visually the histograms are different, it results in a large energy distance.
The permutation test is represented by the lower row, in which the $x,y$ samples have been randomly re-labelled.
In this case it is impossible to distinguish the $x, y$ samples and the resulting energy distance is low.
With many permutations one may construct the grey histogram.

\subsection{Large sample sizes}\label{sec:largesamples}

The computational complexity of \PTED largely comes from computing the distance matrix making it $\mathcal{O}(n^2\, d)$, where $n=n_x+n_y$ and $d$ is the number of dimensions.
For very large samples this quadratic scaling may become untenable and so an option is included to convert the algorithm into a linear scaling method for some loss of sensitivity.
Specifically, the user may select a number of landmark points $m$ and then a rectangular $n\times m$ distance matrix will be computed instead.
This turns the computational complexity into $\mathcal{O}\left(n\, m\, d\right)$ which scales linearly with the sample size.

Restricting to a rectangular distance matrix does not cost exactness, provided the permutations are drawn correctly. 
Let $L \subset \{1,\dots,n\}$ with $|L| = m$ index the landmarks. 
Permutations are drawn not from $S_n$ but from the subgroup $S_L$ and its compliment $S_{L^c}$.
So labels are exchanged within the landmark set and within its complement, but never between them. 
Under $H_0$ the pooled sample is exchangeable, and exchangeable under this subgroup. 
\Equ{permp} therefore retains $\Pr_{H_0}(\hat{p} \le \alpha) \le \alpha$: what shrinks with $m$ is power, not validity. 
Also, the expectation values then take on slightly different normalization; as the full distance matrix is not accessible, it was ultimately easier to compute the expectation values while excluding self-pairs (the all zero diagonal of the distance matrix).
For a discussion on the trade-off between runtime and sensitivity, see \Sec{runtime}.

The landmark elements play a role analogous to the reference points of PQM.
While PQM uses its landmark points to determine a Voronoi tessellation, \PTED directly uses the distances from the landmarks to the rest of the sample.
The same distance matrix is constructed, but each method applies a different interpretation.
An avenue for future exploration may be alternate formulations that retain the useful properties/guarantees of the energy distance.

\subsection{The single-element limit (coverage testing)}\label{sec:singleelement}

An important special case for two-sample tests occurs when $n_x=1$ and $n_y$ may have arbitrarily many elements.
This special case occurs in a number of important contexts including: outlier detection, posterior coverage testing, and posterior predictive checks\footnote{This also applies to frequentest coverage testing, though the discussion here focuses on Bayesian methods~\citep{jaynes1996probability}.}.
Coverage testing is treated by \citet{Cook2006} and \citet{Talts2018} among others; simulation-based calibration (SBC) is the closely related procedure of \citet{Talts2018}.
\citet{Harrison2015} propose a coverage testing method using Highest Posterior Density (HPD) regions and a KS-test.
However, coverage testing is of broad interest in Bayesian Inference as can be seen in \Fig{inference}.

It is possible to apply \PTED in such a single element context.
The single-element case admits a closed form that makes both its behaviour and its limitations transparent. 
Take $n_x = 1$ and write the pooled sample as $z = \{x\} \cup y$.
Substituting $n_x = 1$ into \Equ{samplestat}, the within-$x$ sum is empty and the within-$y$ sum may be written as the pooled total less the terms involving the singleton, giving
\begin{equation}
    \hat{D}^2 = \frac{2\,\delta(z_i)}{n_y-1} - \frac{S(z)}{n_y(n_y-1)},
    \qquad \delta(z_i) = \sum_{j \neq i} \lVert z_i - z_j \rVert,
    \qquad S(z) = \sum_{j}\sum_{j' \neq j} \lVert z_j - z_{j'} \rVert ,
    \label{equ:singleelement}
\end{equation}
\noindent where $S(z)$ is a constant for the purpose of the test. 
The statistic is thus a strictly increasing function of $\delta$, and the permutation test of \Equ{permp} reduces exactly to a rank test on $\delta$:
\begin{equation}
    \hat{p} = \frac{1 + \#\{j \le n_y : \delta(y_j) \ge \delta(x)\}}{n_y+1} .
    \label{equ:singlep}
\end{equation}
\noindent Three consequences follow. 
First, the test needs no permutations at all in this regime; there are only $n_y+1$ configurations in \Equ{singlep}. 
Second, $\hat{p} \ge 1/(n_y+1)$: the number of elements in the second sample sets a floor on the lowest possible $p$-value, irrespective of how far out $x$ lies. 
A single trial with $n_y = 128$ posterior samples can therefore never report $p < 1/129 \approx 0.008$, which is why coverage testing combines many trials (\Sec{implementation}). 
Third, even as $n_y \to \infty$ the statistic remains a single rank, uniform under $H_0$, so power against a fixed alternative is bounded: consistency of \PTED requires $n_x \to \infty$ as well.

When used in the context of coverage testing, the user supplies multiple coverage tests.
Each test compares the single ground truth element with a set of posterior samples, which is the case considered here; this produces multiple independent $p$-values.
In \PTED $p$-values are combined using the Fisher method~\citep{edwards2005ra}. 
Under the null, a $p$-value should be drawn from $U(0,1)$ and it is also the case that $-2\log(p),~p\sim U(0,1)$ is $\chi^2$ distributed with two degrees of freedom. 
The sum of $\chi^2$ values gives another $\chi^2$ distribution with $2n_{\rm sim}$ degrees of freedom from which to apply the coverage test. 
Another $p$-value is computed assuming $-2\sum_i^{n_{\rm sim}} \log(p_i)$ is $\chi^2$ distributed with $2n_{\rm sim}$ degrees of freedom.
A two-tailed $p$-value is computed by considering both sides of the $\chi^2$ distribution, allowing the test to detect overconfidence and under-confidence in the posteriors.

\subsection{Containment testing}\label{sec:containment}

A distinct question from two-sample testing discussed so far, is whether the support of F is contained in that of G.
Rather than checking for $F=G$, one can look for ``$F\subseteq G$'' in some sense.
This question does arise in inference, such as the prior predictive check in step 4 of \Fig{inference} which is more interested in containment than equality.
Containment may have a number of valid/useful definitions in this context: that $G$ has positive density anywhere $F$ has positive density, a bounded total variation distance, or a number of other reasonable choices.
\PTED adopts one such containment test using an approach largely similar to the coverage test in \Sec{singleelement}.

Concretely, pool all the samples and score each element by its depth $\delta(z_i) = \sum_j^{n_y}\lVert z_i-y_j\rVert$ relative to the $y$-labelled subset, so now each $x_i$ may be rank ordered among the $y_i$. 
Each element of $x$ then has the single-element $p$-value of \Equ{singlep} and these are combined using the Fisher method, $C = -2\sum_i^{n_x}\log(p_i)$, as in the coverage test per-simulation $p$-values.
Because every $p_i$ is computed against the same $y$, the $p_i$ are dependent and not $\chi^2_{2n_x}$ distributed. 
\PTED therefore calibrates by permutation rather than against a $\chi^2$ reference, applying \Equ{permp} with $C$ in place of $\hat{D}^2$. 
Large $C$ is evidence that $x$ reaches outside $y$, so the test is one-tailed.

Note that this test is asymmetric, ${\rm containment}(x,y) \neq {\rm containment}(y,x)$ as the first sample is placed among the second for the rank-ordering.
The null is composite: $H_0$ is that $x$ is no more peripheral than $y$, with $F = G$ on its boundary. 
The test is exact on that boundary and conservative in the interior, so a genuinely contained $x$ rejects less often than the nominal rate.
Still, it is a useful and fast gauge of containment, well suited for the early stages (step 4 in \Fig{inference}) of an analysis pipeline.
To further aid the interpretation, \PTED includes a variation on a PIT plot which shows where the $x$ samples live in the $y$ distribution.
If most of the $x$ are well embedded in the $y$ samples and only a few are out in the $p$-value tail then one may tentatively declare containment. 
However, if a large number of $x$ have exceedingly low $p$-value when judged relative to the $y$ samples, then one may infer that $y$ is poorly covering $x$.

The PIT plot guards against certain failure modes of the Fisher combination method. 
The combined $\chi^2$ is a sum, so consider the case of many well contained $x$ samples and a small population of pronounced outliers.
One would wish for this case to be rejected, but the bulk of $x$ sitting deep inside $y$ accumulates slack that can mask a handful of elements far outside $y$. 
Such outliers are clear in the PIT plot as a large tail at low $p$-value, even if the Fisher combination misses them.

\subsection{Implementation Details}\label{sec:implementation}

While the implementation of \PTED agrees very closely with the method described in \citet{szekely2004testing}, a number of minor design choices were made in developing the method, which are described here.
\begin{enumerate}
    \item The diagonal of the distance matrix (all zeros) is excluded from the energy distance expectation value calculation. The resulting estimator of $D^2$ is unbiased and it simplified the normalization for the rectangular distance matrix version in \Sec{largesamples}. The validity of the permutation test is unchanged.
    \item  $\frac{q+1}{N+1}$ is used to compute a $p$-value from a finite number of permutations, where $q$ is the rank of the test among the permutations and $N$ is the number of permutations. \citet{szekely2004testing} use $q/N$, however this allows invalid $p = 0$ scenarios, and is a less robust estimator for permutation test based $p$-values~\citep{Phipson2016}.
    \item Two-tailed $p$-values are computed with $2\min(p_{\rm left}, p_{\rm right})$, see \Equ{twotailed}.
    \item Computing the energy distance requires evaluating all pairwise distances in the combined two samples $O(n^2\, d)$, this is also true of any permutation. Rather than re-computing the distance matrix, \PTED re-uses the same distance matrix, and uses indicator vector multiplication, $O(n^2)$, to perform the permutation test; which for large $d$ can be an immense speedup.
    \item For the sake of computational speed, the user may supply sample arrays on a GPU (via JAX or PyTorch) for faster calculation of the distance matrix.
\end{enumerate}

\noindent \PTED is available on GitHub at: \href{https://github.com/ConnorStoneAstro/pted}{https://github.com/ConnorStoneAstro/pted} and may be installed on most systems with:
\begin{center}
    \texttt{pip install pted}
\end{center}
\noindent The GitHub also includes a README with instructions on how to call the two-sample test, or the coverage test and how to extract various metrics, including an SBC diagram and PIT plot.

\section{Comparison tests}\label{sec:tests}

Here a number of controlled two-sample test experiments are run to compare the performance of \PTED with other algorithms.
The most straightforward to compare with is the KS-test, although some of the tests extend to multiple dimensions in which case the samples are projected along the first principal component of the data to convert it to one dimension.
The second is PQM~\citep{Lemos2025}, which uses a Voronoi tesselation of the sample space to bin the samples into cells, then uses a multinomial distribution under the null to compute a $p$-value.
Standard practice with PQM is to run many re-tesselations of the samples to median average the $p$-values, which gives a more stable result, though it means the reported $p$-values are no longer $U(0,1)$ distributed.
Third, we compare with MMD in its more common form in machine learning literature which uses the Radial Basis Function (RBF) kernel.
Following \citet{Gretton2006}, the RBF scale is set to the median pairwise distance in the pooled sample.
This median heuristic is a convention rather than an optimal choice.
Fourth is the FID score, widely used to compare generative model samples in high dimensions, though it is a divergence estimate rather than a null hypothesis test.
The FID score is based on a multivariate Gaussian fit to the two samples and then uses the Wasserstein distance between the results.
The fifth method is the FLD score, also often used in high-dimensional two-sample tests while not explicitly being a null hypothesis test~\citep{jiralerspong2023feature}. 
FLD is based on a mixture of Gaussians fit to the samples to compute a likelihood of one sample given another as a reference.
FLD and FID do not produce calibrated quantities and one is encouraged to use relative values as references.

\subsection{1D Gaussian sensitivity tests}\label{sec:gaussian1d}

\PTED and the five comparison methods are compared against one-dimensional normally distributed samples.
For the first sample, $x$, 100 normally distributed points are drawn; for the second sample, $y$, a modified distribution is used.
In each test a severity score $S\in[0,1]$ interpolates the sampling distribution for the $y$ samples from the standard normal distribution ($S=0$) to an alternative distribution ($S=1$).
Two tests are run by modifying the mean and standard deviation of the $y$ samples and applying the two-sample tests (see \App{gaussian1d} for a visualization).
Each configuration is repeated 64 times with independent draws; figures report the median and the 16th--84th percentile range across repetitions, and all permutation tests use 512 permutations.

\begin{figure*}
    \centering
    \includegraphics[width=0.8\linewidth]{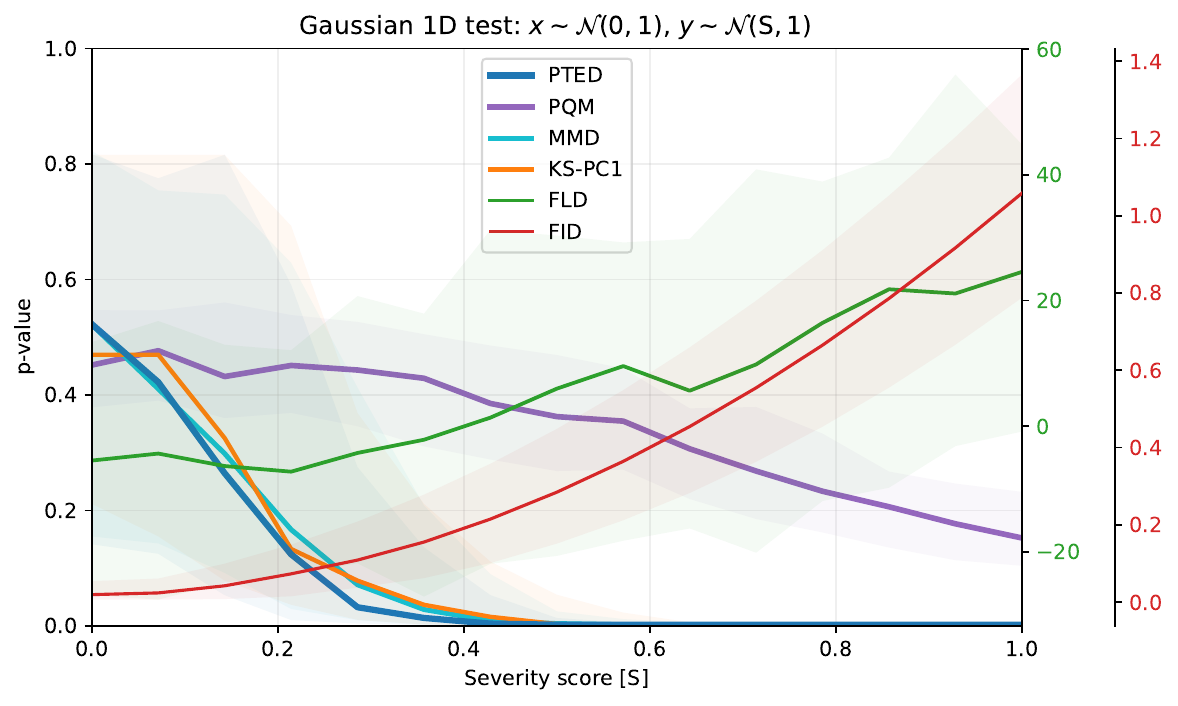}
    \includegraphics[width=0.8\linewidth]{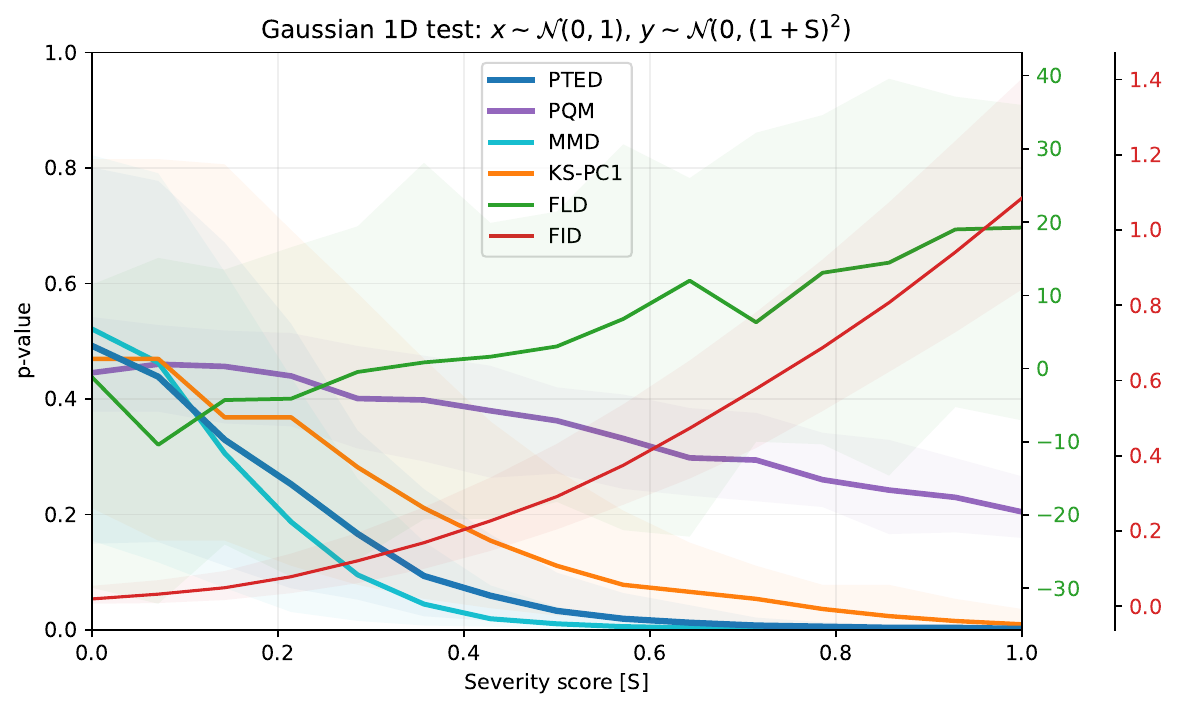}
    \caption{Comparison of two-sample tests with \PTED under 1D Gaussian distributed data. The $x$ samples are 100 points from a normal distribution, the $y$ samples are 100 points from a modified normal distribution as indicated in each sub-figure. For each algorithm, the median is plotted as a thick coloured line and the 16-84 percentiles are shown as a shaded region. Top: the $y$ sample mean is set to $S$ (the ``severity score''). Bottom: the $y$ sample standard deviation is set to $1+S$. }
    \label{fig:gaussian1d}
\end{figure*}

\Fig{gaussian1d} presents the results of these tests as a function of severity score.
In each test, the median \PTED $p$-value is more sensitive than the other $p$-value tests (KS-test and PQM), except MMD which is essentially identical. 
Interpreting FID and FLD is challenging as their output range changes for each test (note right side coloured y-axes). 
The FID score appears highly sensitive to the out of distribution samples, as should be expected given its construction under the assumption of Gaussianity.
A permutation test on the FID score would be worthwhile for future consideration.
The FLD score appears largely insensitive to the deviations, only deviating from the $S=0$ score after considerable deviation from normality.

A further instructive test is not shown: $y\sim\mathcal{N}$ except with a single outlier.
No two sample test was able to detect this failure mode regardless of how far the outlier was set from the rest of the samples.
The reasons for this are related to pure sample testing and are discussed in more detail in \Sec{highdim}.

\subsection{Two-Moons test}\label{sec:twomoons}

In this next test, a hypothetical model fit to the two-moons distribution is considered.
To begin, 256 samples are drawn from the two-moons distribution as training data.
Those 256 points are chosen as the centres for a 256 component Gaussian mixture model, which represents a hypothetical fitting/machine learning algorithm.
For the mixture of Gaussians, a number of standard deviations are considered, from far under-fitting the two-moons, to highly over-fitting the training samples.
Note that a mixture of Gaussians with essentially zero standard deviation (the far over-fitting condition) is the optimal solution for a score matching objective learning function used for diffusion models~\citep{song2020} and so this example is not so contrived as it may seem.

\begin{figure*}
    \centering
    \includegraphics[width=0.205\linewidth, trim={0 0 18cm 0}, clip]{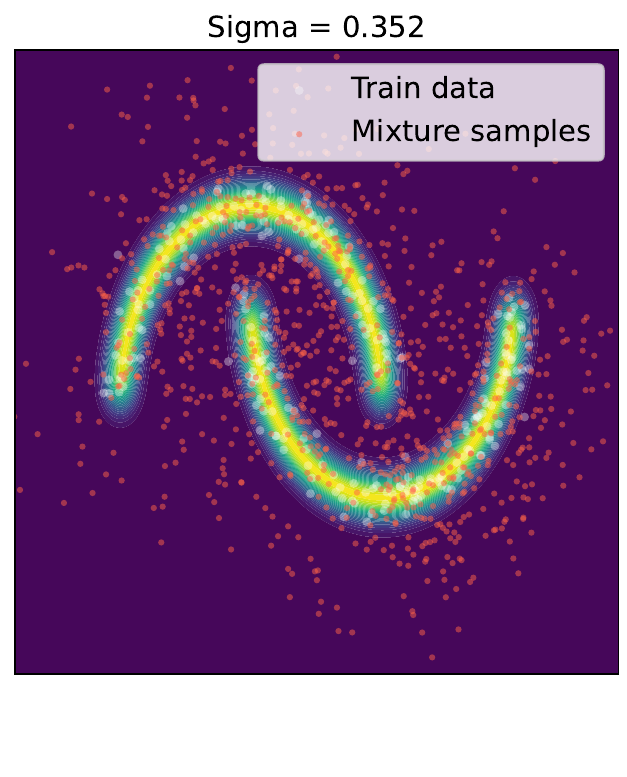}
    \includegraphics[width=0.205\linewidth, trim={0 0 18cm 0}, clip]{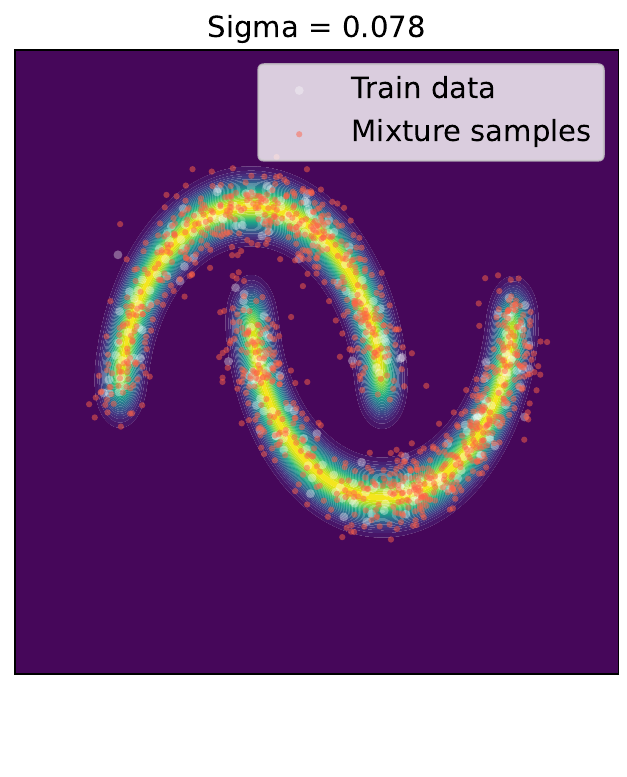}
    \includegraphics[width=0.56\linewidth]{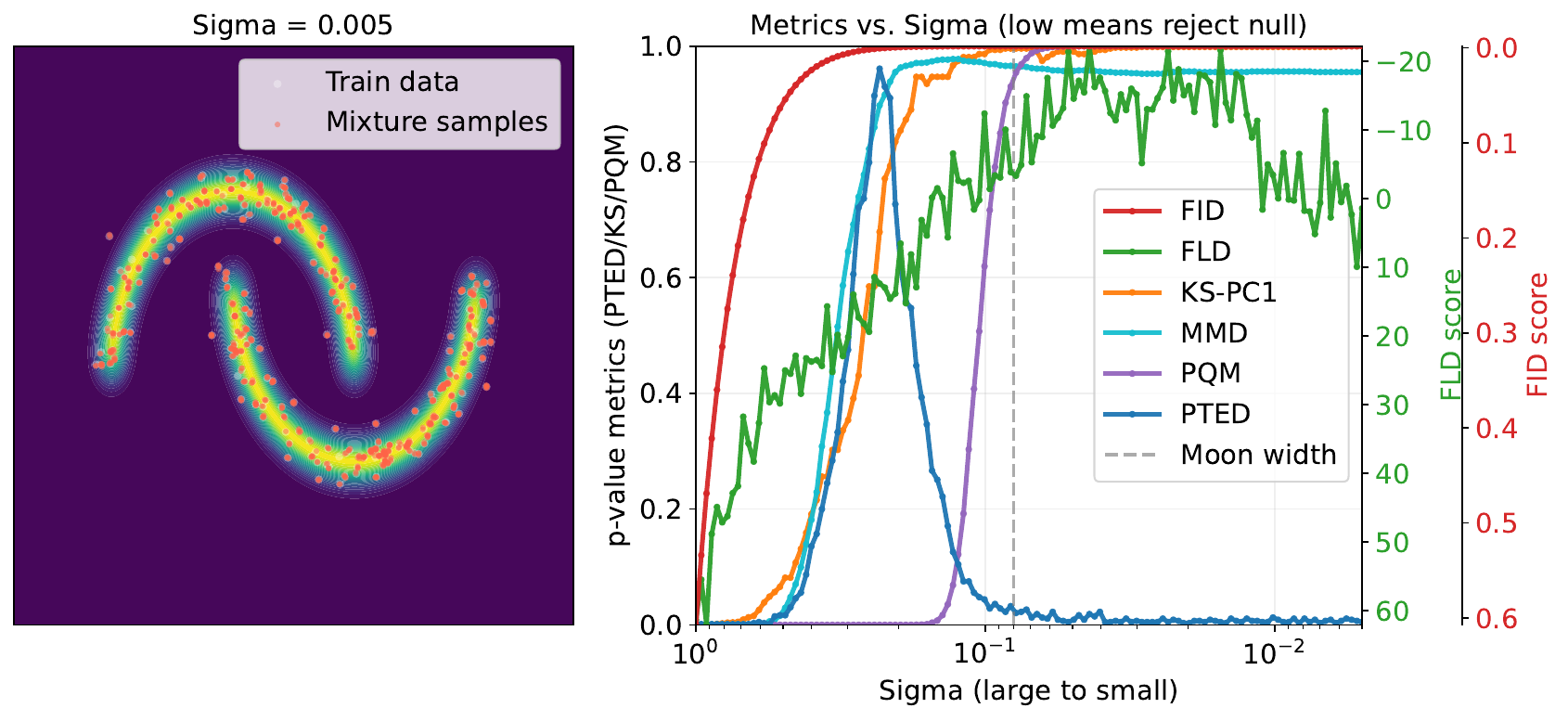}
    \caption{Two-Moons over-fitting progression test. The two-moons distribution is shown in the first three panels with viridis colour scale for the density. White points are samples from the two-moons distribution, red points are from the hypothetical model trained on those data. The first three panels show a progression of the Gaussian mixture model standard deviation, from far under-fit on the left to far over-fit on the right. The rightmost sub-figure gives the progression of each metric when sweeping from highly under-fit to highly over-fit to the training data. The FLD and FID metrics have been flipped on the y-axis so that all metrics show a ``higher is better'' evaluation of the fitting progression.}
    \label{fig:twomoons}
\end{figure*}

In \Fig{twomoons} the results of this experiment are presented.
Three panels of the figure show the two-moons distribution alongside samples from the mixture of Gaussians with different scales. 
The first panel shows clear under-fitting, and yet FID, MMD, KS-test, and \PTED all fail to reject the null.
Such blind-spots are expected as no two-sample test is universally most sensitive.
PQM has its $p$-value spike near the 0.08 value used as the width of the two moons; while there is no particular scale that would allow the mixture of Gaussians to exactly reproduce the two-moons distribution, selecting a similar scale is a reasonable choice.
FLD continues to rise even past the 0.08 scale and then fall again when it is clear the samples are now over-fitting the training data.
It might be difficult to interpret in a real training scenario as the FLD score has no inherent meaning, but in this test the FLD score picks out both good fitting and over-fitting.
It is apparent in the \PTED $p$-values that it detects over-fitting as well given the steep drop in $p$-value below the 0.08 scale.
This is because \PTED can operate in a two-tailed mode which is sensitive to under-fitting and over-fitting.
MMD can also operate in a two-tailed mode which would have detected over-fitting as well, but this is not commonly how MMD is used.

\subsection{MNIST sensitivity tests}\label{sec:mnist}

The tests considered in \Sec{gaussian1d} and \Sec{twomoons} operate in one and two dimensions respectively, however some of the most exciting applications of these two-sample tests are in many hundreds or thousands of dimensions.
Here a number of tests are run on the classic MNIST dataset~\citep{lecun1998gradient}.
This dataset is composed of $28\times 28$ pixel images of handwritten digits from zero to nine, it is commonly used as the simplest test case for machine learning algorithms.
Three tests are considered: adding white noise, progressively removing the zero digits, and randomly blending pairs of digits (see \App{mnist} for a visualization).
Each of these deviations represents a failure mode for generative models.
White noise represents noise that can be added by incomplete training failing to land exactly on the low-dimensional manifold of the digits embedded in the $28^2$ dimensional space.
Dropping one of the classes can occur in generative models which, for some reason or another, discover a short cut to minimize their loss by focusing on a subset of the data classes.
Blending samples can occur in regression models that learn some mean trend rather than attempting to represent the probability distribution from which the samples are drawn.
Both $x$ and $y$ samples contain 2048 elements to test sensitivity with intermediate size samples.
Increasing the sample sizes will make all methods more sensitive.
In practice while training a generative model and using these tests for validation, the practitioner would need to balance generating enough samples to detect an issue with the runtime costs of generating many samples.

\begin{figure*}
    \centering
    \includegraphics[width=0.75\linewidth]{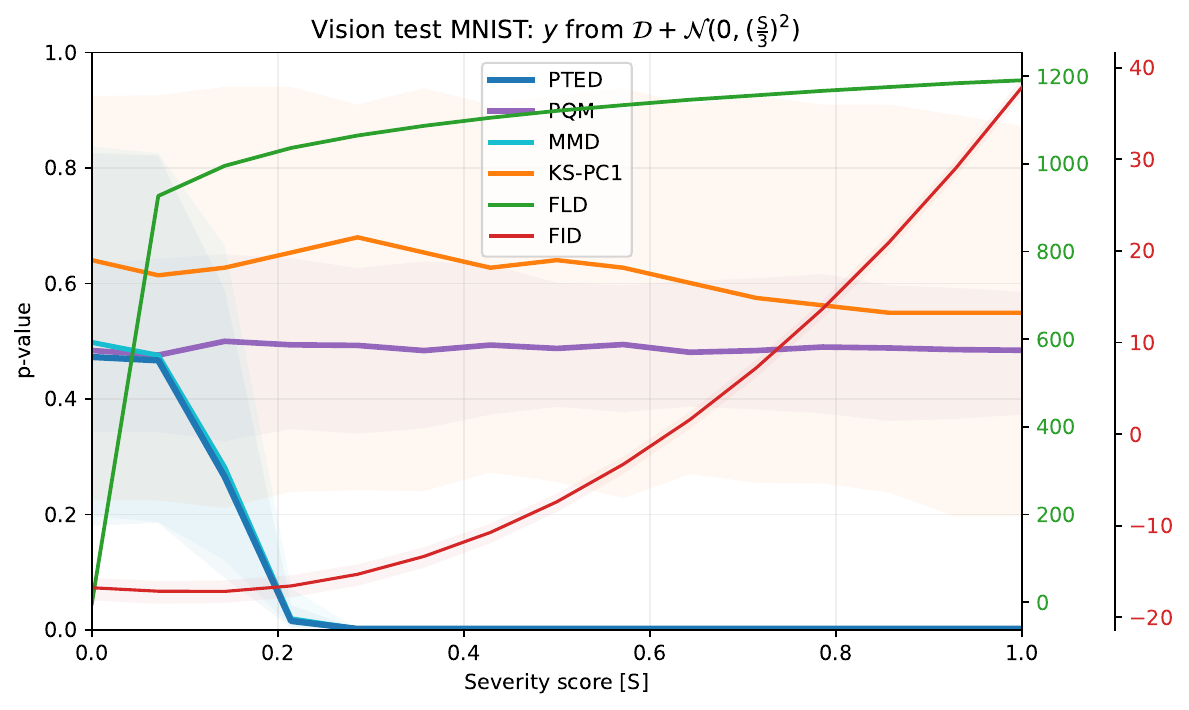}
    \includegraphics[width=0.75\linewidth]{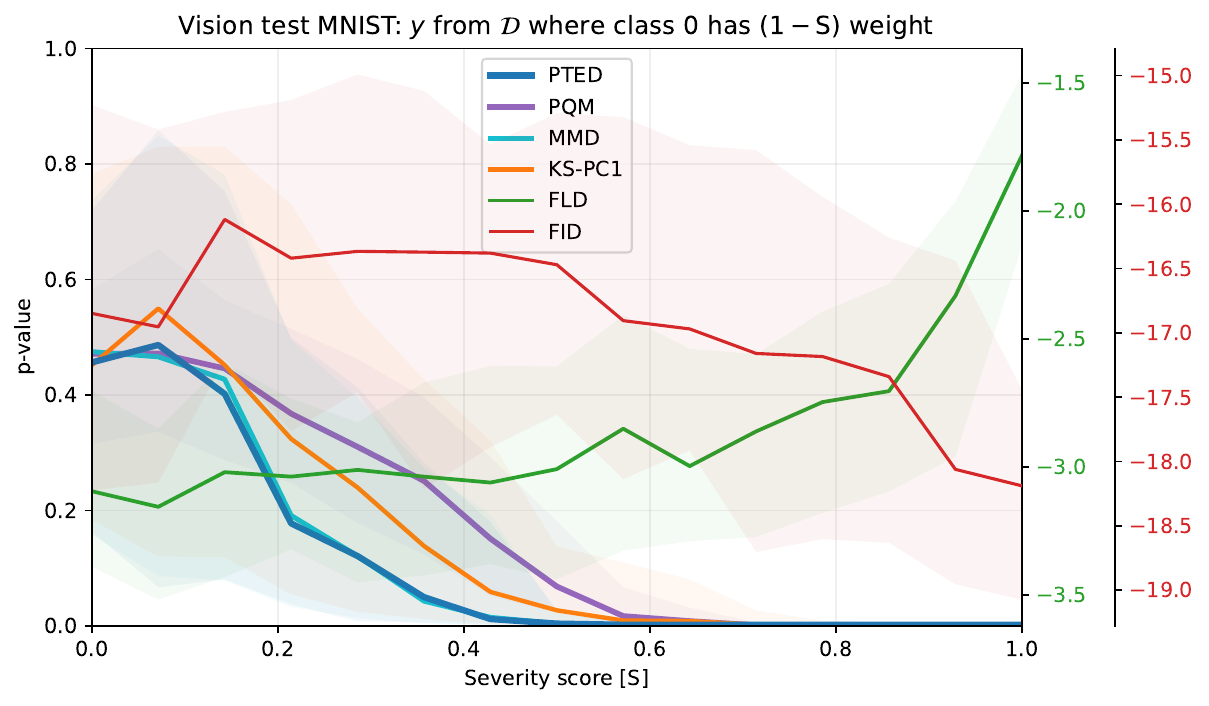}
    \includegraphics[width=0.75\linewidth]{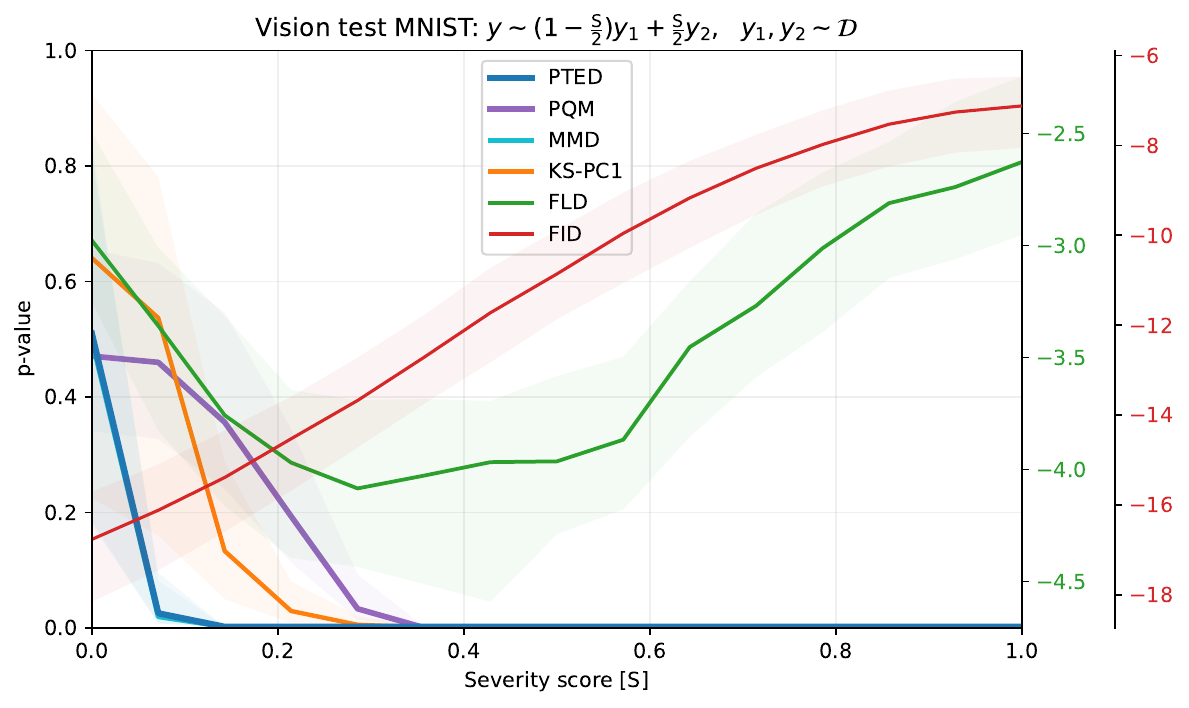}
    \caption{Comparison of two-sample tests with \PTED under MNIST data distribution. The $x$ samples are 2048 MNIST images drawn from the full sample, the $y$ samples are 2048 MNIST images drawn from the full sample (less the $x$ samples) plus a modification as described for each distribution. Top left: the $y$ samples have a small amount of white Gaussian noise added to them. Top right: the $y$ samples have a reduced weight for drawing a zero MNIST digit. Bottom: the $y$ samples are composed of two random MNIST digits with progressively more significant mixing. Note: many FLD runs returned NaN results (likely due to the low variance in some MNIST dimensions), the FLD lines (green) include only those tests that returned finite values.}
    \label{fig:mnist}
\end{figure*}

\Fig{mnist} presents the comparison tests applied to MNIST data for each two-sample test metric.
\PTED and MMD have nearly identical performance in all the tests, this is somewhat surprising given their entirely different kernels.
\PTED uses the Euclidean distance directly and so increases as two elements move apart, while MMD uses an RBF kernel which decreases as a function of Euclidean distance. 
The KS-test and PQM are entirely insensitive to white noise, while FLD appears more sensitive than \PTED.
Neither FID or FLD appear sensitive to the class drop test.
The FLD metric decreases slightly for the pair blending test, suggesting it favours the more severely distorted samples; this is the opposite of its intended behaviour.
This result is replicated in the CIFAR-10 tests in \App{cifar10}.

\subsection{Coverage Test}\label{sec:coverage}

As discussed in \Sec{singleelement}, it is possible to run \PTED in a mode suitable for Bayesian posterior coverage testing.
This tests a different category of problem since one of the samples has only a single element (the ground truth).
Only \PTED, MMD, and the KS-test are able to operate in this regime; FLD, FID, and PQM require enough elements in each sample to fit or bin and so are excluded for this section.
To allow for further comparison, an HPD region coverage test and the MIRA algorithm are included.
The HPD region test first rank-orders the ground truth among the posterior samples via the posterior density, computes a $p$-value using $\frac{q+1}{N+1}$, then combines multiple simulations using Fisher's method.
This is identical to \PTED in \Sec{singleelement}, except using the density for rank-ordering rather than the energy distance.
MIRA follows PQM in exploiting the observation that regions of equal posterior mass should contain comparable numbers of samples~\citep{Sharief2026}.
MIRA does not output a $p$-value, but instead returns a score value that lands within a known interval 68\% of the time when the null is true.

\begin{figure}
    \centering
    \includegraphics[width=0.8\linewidth, trim={0 12cm 0 0}, clip]{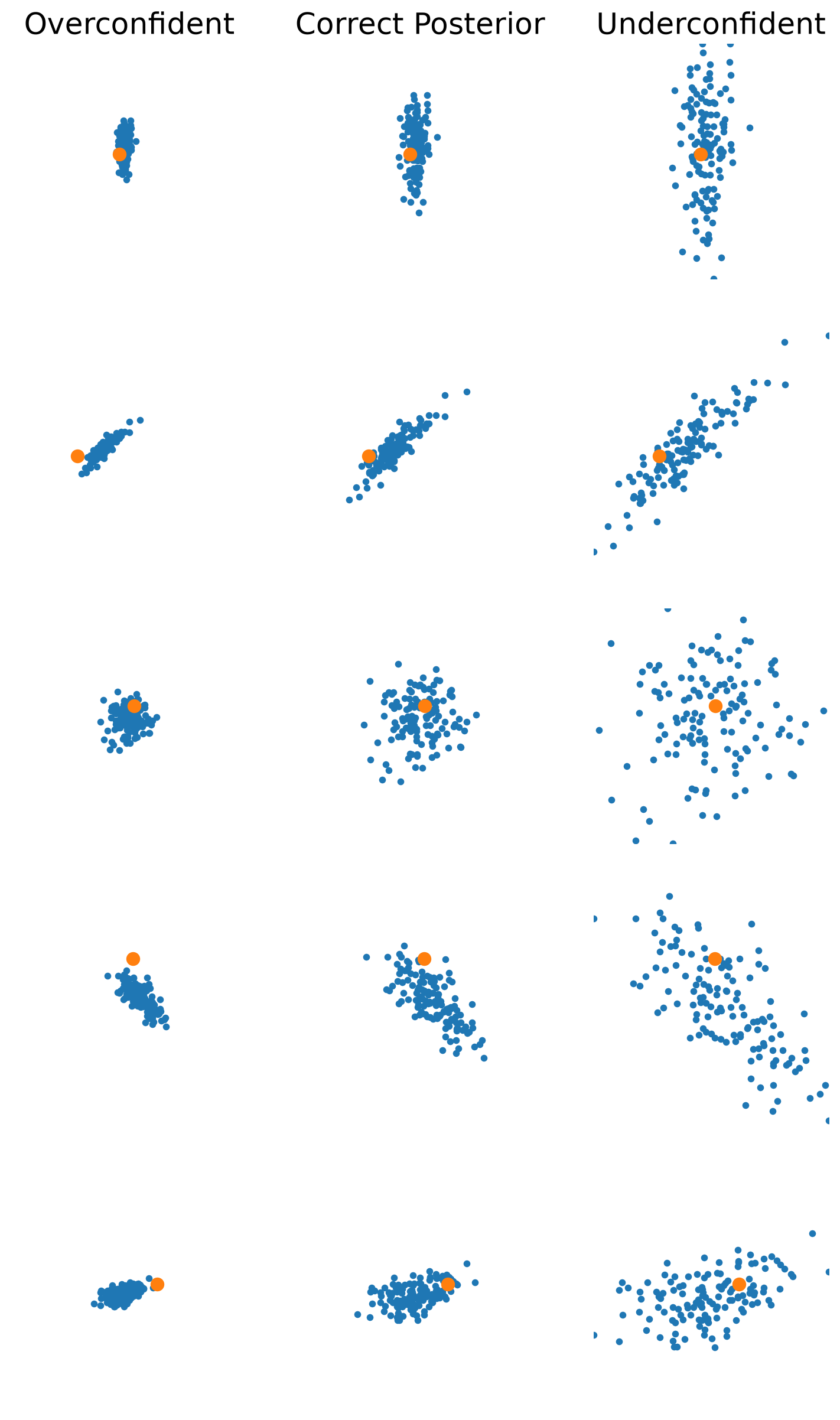}
    \caption{Example scaled posterior sample distributions (blue) compared with their respective ground truths (orange). Left column: over-confident examples, the posterior distribution covariance matrix is scaled by $\frac{1}{2}$ producing overly narrow posteriors that do not consistently cover the ground truth. Centre column: correctly calibrated posterior samples, the ground truth is indistinguishable from a posterior sample. Right column: under-confident example posteriors, the posterior distribution covariance matrix is scaled by $2$ producing overly wide posteriors making the ground truth overly centralized in the posterior distribution.}
    \label{fig:coverageexamples}
\end{figure}

Posteriors in many inference problems are well approximated by a Gaussian, so the experiment uses Gaussian mocks.
For $n_{\rm sim} = 64$ simulations, a random $2\times 2$ covariance matrix $\Sigma$ is drawn: $\sigma_x,\sigma_y\sim U(0.5,2)$ and $\sigma_{x,y}^2 \sim U(-\sigma_x\sigma_y, \sigma_x\sigma_y)$ representing the true posterior.
A single sample is drawn to represent the ground truth, and then 128 samples are drawn to represent the posterior.
In \Fig{coverageexamples} some example posterior distributions and their corresponding ground truths are displayed.
To test the various algorithms, the posterior covariance is scaled as $c\Sigma$ for $c\in(0.5,2)$ while the ground truth remains fixed to produce over-, under-confident results.
When the posterior samples are scaled smaller, this effectively makes the ground truth appear as an outlier among the posterior samples.
When the posterior samples are scaled larger than the ideal, this makes the ground truth appear near the central regions of the posterior samples more often than should be expected, suggesting under-confidence.
This effect is more subtle and can only be identified as a discrepancy over many simulations.

\begin{figure}
    \centering
    \includegraphics[width=\linewidth]{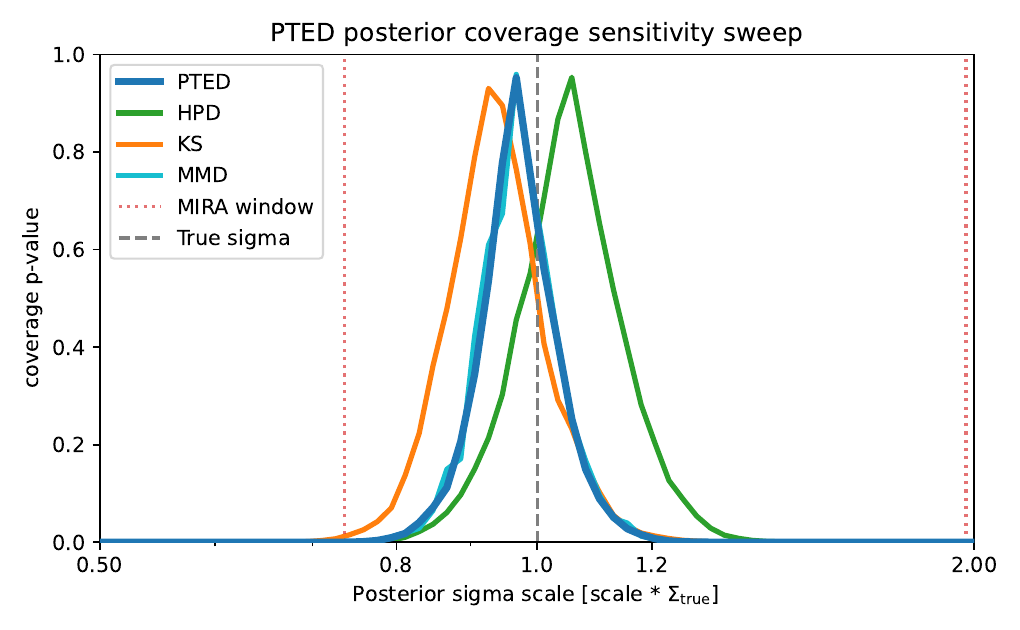}
    \caption{Coverage $p$-values for \PTED, the KS test and the HPD-region test, as a function of the factor $c$ by which the posterior covariance is rescaled (\Fig{coverageexamples} shows examples). MIRA returns a score rather than a $p$-value, so it is shown differently: the vertical dotted lines bound the range of $c$ over which MIRA's score remains inside its 68\% null interval, i.e.\ the range it does not flag.}
    \label{fig:coveragetest2d}
\end{figure}

\Fig{coveragetest2d} presents the resulting $p$-values for each method applied to the mock simulated datasets.
Since the test is done in two dimensions and the KS-test only operates in one dimension, values of the two dimensions are summed (via a dot product with a vector of ones) to produce a single value.
All the considered algorithms are able to identify both over- and under-confident results with varying degrees of sensitivity.
MIRA is the least sensitive: its null window extends to nearly $c=2$ while the other methods reach the same significance by $c\approx 1.3$.
The KS-test and HPD region appear similarly sensitive but with peaks on opposite sides of the ideal value.
\PTED and MMD give the highest sensitivity of all the methods, though only by a small margin on either over- or under-confidence.

\section{Discussion and Conclusions}\label{sec:conclusion}

\subsection{Sensitivity across the test suite}

\begin{figure}
    \centering
    \includegraphics[width=\linewidth]{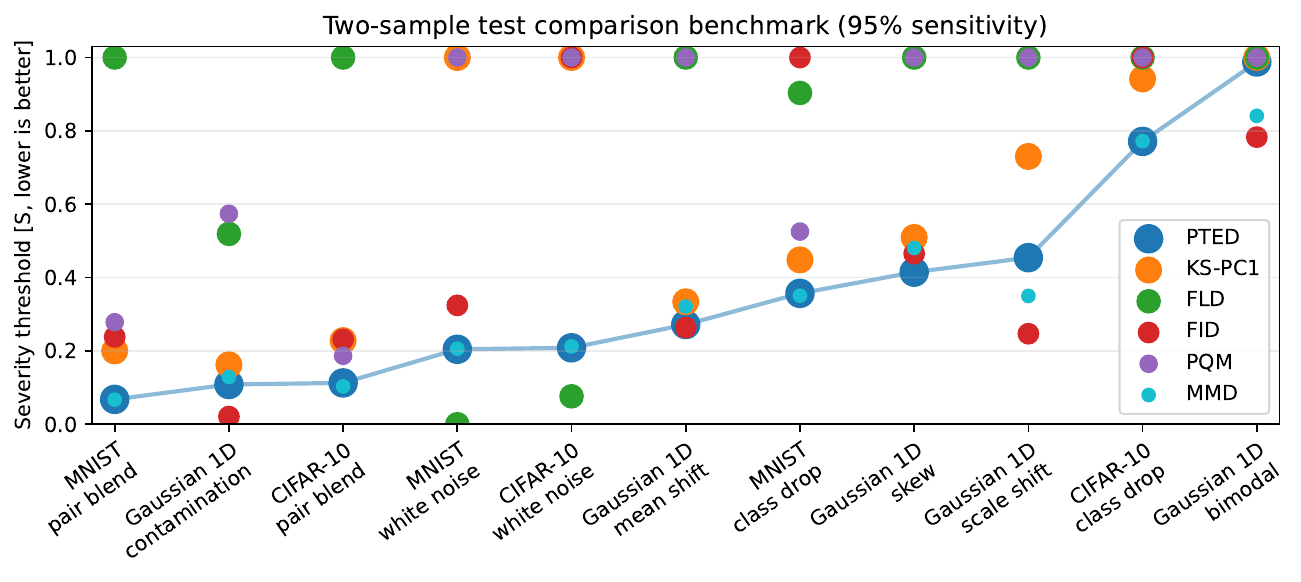}
    \caption{
    Combined test results from pure two-sample test comparisons in \Sec{gaussian1d}, \Sec{mnist}, \App{gaussian1dextra}, \App{cifar10}. For each method, the plotted value is the smallest severity score $S$ at which the median $p$-value over realizations falls below $0.05$. FID and FLD do not produce $p$-values, so for these the plotted value is the smallest $S$ at which the median score falls outside the central 95\% range of scores obtained at $S=0$. A value of one indicates that the method did not reject at any severity. The tests are sorted by \PTED performance, connected by the blue line for visual guidance. A severity score below the line indicates superior performance to \PTED.}
    \label{fig:benchmark}
\end{figure}

\PTED consistently performed well across the full range of two-sample test problems presented.
Of all the algorithms, it was the only one to produce competitive sensitivity on all tests as can be seen in \Fig{benchmark}.
The FID test showed impressive sensitivity in a number of tests, but failed to detect replication in the two-moons test (\Sec{twomoons}), and was insensitive to the class drop test in MNIST (\Sec{mnist}) and the white noise test in CIFAR-10 (\App{cifar10}).
FLD managed to detect replication in the two-moons test, but was largely insensitive to a number of other perturbations (e.g., all Gaussian 1D tests, and MNIST/CIFAR-10 pair blending).
The KS-test was reliably nearly as sensitive as \PTED, but at no point was more sensitive, failed to detect replication in the two moons test, and in the MNIST/CIFAR-10 white noise tests failed to detect even the most severe deviation.
PQM showed a mix of competitive sensitivity, and complete insensitivity to a number of tests.
MMD, was essentially identical to \PTED in all tests; though it failed to detect replication in the two-moons test this is merely a detail of the typical implementation used in generative modelling contexts.

\Fig{benchmark} suggests a bimodal pattern: each method either shows similar sensitivity to the rest, or is unable to detect the deviation at all.
If this trend generalizes, then consistency matters more than slight sensitivity gains.
The fact that \PTED (and MMD) are provably sensitive to any deviation, given enough samples, is compelling in light of this bimodal trend.
While it may not be the most sensitive in a fine grain test such as performed here, for real-world applications it is often the case that such minor performance differences are irrelevant.
The fidelity, diversity, and novelty criteria used to assess generative models are, jointly, a restatement of the two-sample problem: a sample that is indistinguishable from the target distribution satisfies all three. 
A calibrated multivariate two-sample test therefore addresses them by construction.
By directly presenting a multi-dimensional two-sample test (e.g., \PTED) such concerns are solved in general and by construction.

\begin{table}[]
    \centering
    \begin{tabular}{r|c c c}
        Method & 1D Gaussian & MNIST & CIFAR-10  \\\hline
        \PTED & 0.001 & 0.008 & 0.03 \\
        KS-test & 0.001 & 0.07 & 1 \\
        MMD & 0.09 & 0.4 & 0.4 \\
        FLD & 2 & 31 & 31 \\
        FID & 0.0006 & 0.8 & 1 \\
        PQM & 0.4 & 0.5 & 0.8
    \end{tabular}
    \caption{Runtime comparison on tests in \Sec{tests}. All runtimes are in seconds. The 1D Gaussian test had one dimension and 100 samples. The MNIST test had $28^2$ dimensions and 2048 samples. The CIFAR-10 test had $3\times 32^2$ dimensions and 2048 samples. Tests were run with compute heavy elements on a GPU (NVIDIA RTX A5000) and other elements on a single CPU core (AMD Ryzen Threadripper PRO 3995WX 64-Cores). Runtimes varied by $\sim 10\%$ from test to test.}
    \label{tab:runtimes}
\end{table}

\subsection{Runtime and scaling}\label{sec:runtime}

Runtimes for each method varied considerably among the tests in \Sec{tests}.
For the sake of comparison, average runtimes are included for a number of tests from \Sec{tests} in \Tab{runtimes}.
The KS-test and FID scaled the most steeply with dimensions, being the fastest for one-dimensional Gaussian data, but for the high-dimensional image data became considerably slower.
This is almost certainly related to the large matrix operations they used.
FLD slowed considerably for the high-dimensional tests, and was slowest overall, likely due to the fitting step internal to the algorithm.
PQM achieved intermediate speeds; individual iterations (re-tesselations) are fast, but running 512 iterations of the algorithm slowed it down.
MMD and \PTED are mathematically near identical and would perform at similar speeds, except that for MMD a simpler permutation method was used involving permuting the rows and columns of the distance matrix to perform the permutation test (for discussion on \PTED implementation see \Sec{implementation}).
\PTED was among the fastest algorithms in each test, with the lowest runtime for the CIFAR-10 test.
These rankings could change as the number of dimensions or samples are adjusted.
Given that each algorithm has different scaling with dimensions and number of samples, it is not surprising that no clear fastest algorithm emerged.
Further, the GPU likely did not saturate for these examples and so the flat performance of PQM is not truly showing its runtime scaling.

\begin{figure}
    \centering
    \includegraphics[width=0.51\linewidth]{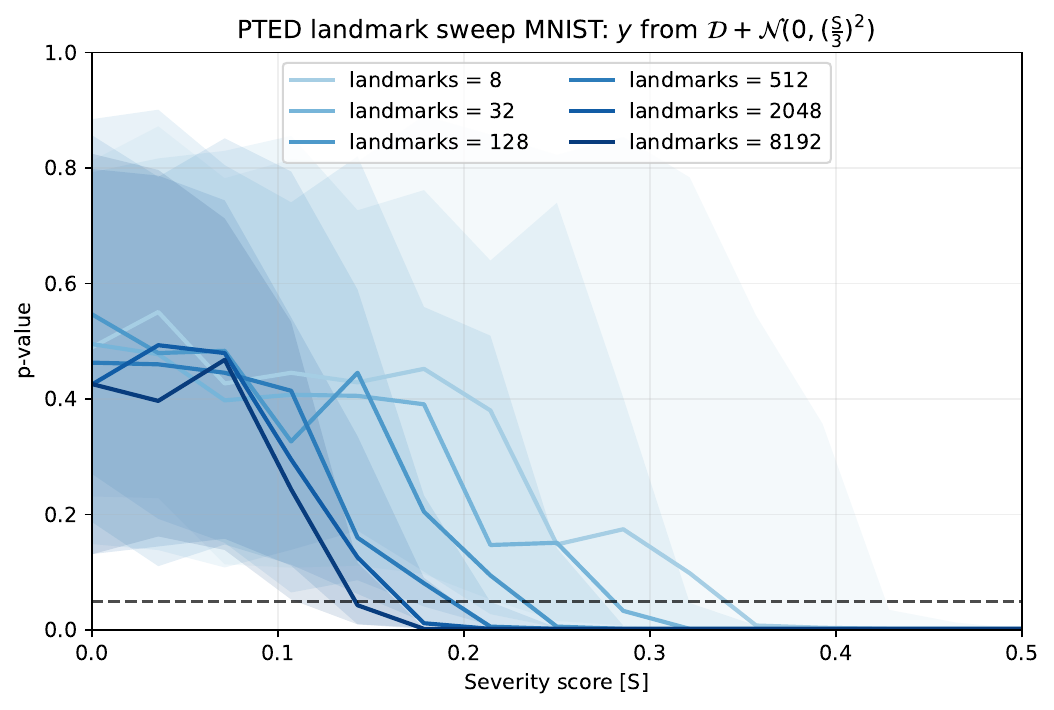}
    \includegraphics[width=0.47\linewidth]{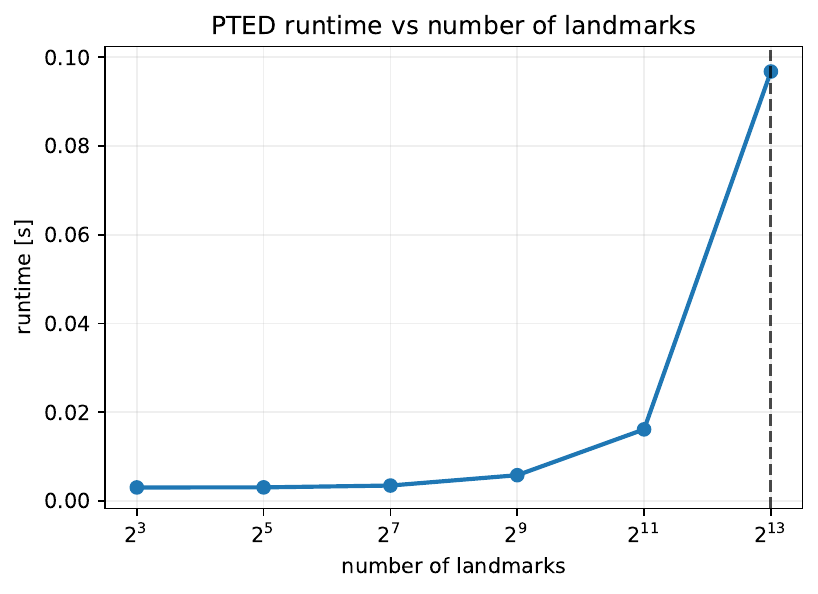}
    \caption{Visualization of the \PTED landmark system trade-off of runtime and performance. Left: performance on the MNIST white noise test from \Sec{mnist} as the number of landmarks is reduced from 8192 (full distance matrix) to 8. A grey horizontal dashed line shows a $p=0.05$ null rejection threshold. Right: the average runtime of 32 runs for each landmark size. A vertical grey dashed line shows the landmark size for which the full dataset (standard \PTED) is used. A larger number of samples (8192 compared to 2048 in \Sec{mnist}) are used for the runtime test to better show the thinning performance.}
    \label{fig:landmark}
\end{figure}

With the ``landmark'' scheme implemented for \PTED, it can operate in a mode that scales linearly with number of samples.
All other methods (except standard MMD) also scale linearly with the number of samples.
\Fig{landmark} presents the trade-off in discriminative power and runtime performance with the landmark scheme.
A near order of magnitude speedup can be achieved with almost no loss in power by using one sixteenth of the samples as landmark samples, cutting the distance matrix from $8192\times 8192$ to just $8192\times 512$.
Even reducing the number of landmarks to 8 (a factor of a thousand reduction), the test rejects the null at a severity of $0.35$ (comparable to FID).

\subsection{Practical considerations and high dimensions}\label{sec:highdim}

One test not shown in \Sec{gaussian1d} is the case of $y$ sampled from a Normal distribution, except with a single outlier.
Essentially, this test considers a well-matched $y$ sample except for some single ``catastrophic failure'' which can occur in real world scenarios due to numerical instabilities or incomplete network training.
No test considered here rejected the null in this configuration, however extreme the outlier. 
This is a structural property of sample-based tests rather than a deficiency of any one of them.
While initially surprising, it is ultimately a consequence of purely sample-based tests.
\PTED and the KS-test are the simplest to explain, as exact tests they are correctly calibrated for \emph{any} sampling distribution, this includes distributions with some extra small mode far off in parameter space.
With only a single far outlier example, the data is indistinguishable from the null under the case that the $x$ samples happened to have not sampled from the small far off mode.
However, if one were to increase the sample sizes and show that many $y$ samples came from such far off outlier modes while $x$ did not, or if the $y$ sample was very small (say two elements with one being the far outlier), then the probability under the null can become vanishingly small.
This is instructive for purely sample-based tests, they must allow for a great deal of flexibility/leeway to truly produce valid results on arbitrary sampling distributions.

There remain a number of practical considerations for real high-dimensional tests.
Most notably, for many high-dimensional generative models, it is known that the density distribution of interest exists as a low-dimensional manifold embedded in the full high-dimensional space~\citep[the ``Manifold Hypothesis'', see][]{fefferman2016testing}.
A perturbation that is small in Euclidean norm but directed off the manifold may be obviously out of distribution to the eye while leaving pairwise distances almost unchanged, and hence be nearly invisible to a distance-based test.
This is not a failure of two-sample tests, or of distance-based methods, but rather the formulation of the problem.
In high dimensions many coordinates contribute to the statistic. This aggregation is what makes the test powerful, but it also means that a discrepancy confined to a few coordinates is diluted by the variation in the rest\footnote{In some sense, high-dimensional two-sample tests suffer from ``multiple hypothesis testing'' dilution of sensitivity. A deviation that would be definitive in a single dimension, may occur often even under the null in high dimensions.}.
Even a large discrepancy on a single axis (feature/dimension/etc.) may be written off as a fluke for the purpose of two-sample testing given only sample-based information (see the discussion of single outliers in \Sec{gaussian1d}).
This is because it is rare or impossible to densely fill a high-dimensional space, and so the low-dimensional manifold of the generating distribution is not fully borne out.
Thus the common practice in machine learning to operate on feature spaces, typically a feature layer in some neural network trained on similar (or the same) data to that being tested.
By projecting into the low-dimensional manifold, each dimension becomes far more informative and a discrepancy in this space may be treated more powerfully as possible evidence to reject the null.
The trade-off is that the projection into a lower dimensional manifold may obscure a failure mode that is orthogonal to the projection.

High-dimensional two-sample tests can be powerful, as demonstrated in the MNIST tests in \Sec{mnist} even a small amount of noise can move the samples measurably out of distribution.
For some problems, passing a \PTED test may be a higher bar than any current generative model can achieve.
In these cases, \PTED is still a useful tool as relative comparisons of the $p$-values or even the energy distances directly can rank-order a series of models.
The $p$-value statistic in \PTED is based on rank-ordering, so more extreme $p$-values imply more extreme differences between generated samples and a test set.
Further, it is worth noting that the energy distance is a differentiable metric, an avenue for further exploration might be to train on the energy distance itself.

\subsection{Conclusions}

\PTED is a versatile and sensitive two-sample test.
It works in low or high dimensions, with small or large samples, with imbalanced samples, and for any underlying distribution it is always an exact test.
For much of the Bayesian inference pipeline, machine learning generative model testing, and SBI validation, \PTED presents a powerful quantitative and interpretable check that is invaluable in guiding progress.
\PTED includes a number of pre-built tests that are well suited to the inference workflow: the base two-sample test, coverage test, and containment test.
This makes it a near universally applicable two-sample testing tool, though it may not always be the most sensitive.

\section*{Acknowledgements}

CS thanks Ren{\'e}e Hlo{\v z}ek for her insightful comments that turned this manuscript from a series of facts into a story.
CS thanks Alex Malz for comments that improved the presentation and rigour of the manuscript.
CS thanks Fangyi Zhu for adding a TQDM progress bar to the \texttt{Python} implementation of \PTED.
CS thanks Sammy Sharief, Yashar Hezaveh, Laurence Perreault-Levasseur, Alexandre Adam, Ronan Legin, and Gabriel Missael Barco for insightful discussions on Bayesian Inference and the power of Sample Based Inference. 

During the preparation of this manuscript CS used Anthropic Claude (Opus 5, accessed September 2026) to improve the clarity of the presentation.
Anthropic Claude was also used to improve the performance of the \PTED software implementation.
During the preparation of this manuscript CS used GitHub Copilot (GitHub Inc., accessed August 2026) to construct the testing framework that produced most figures.
See \url{https://github.com/ConnorStoneAstro/pted_tests} for a public repository to reproduce all the tests and figures (except \Fig{inference} and \Fig{ptedmethod}).

\section*{Software}
\PTED~\citep{ptedzenodo}, numpy~\citep{numpy}, PyTorch~\citep{pytorch}, JAX~\citep{jax}, SciPy~\citep{scipy}, Matplotlib~\citep{matplotlib}, FLD~\citep{jiralerspong2023feature}, PQM~\citep{Lemos2025}, MIRA~\citep{Sharief2026}, TQDM~\citep{casper2022}, Python~\citep{python}

\bibliographystyle{unsrtnat}
\bibliography{pted}

\appendix

\section{Extra Gaussian sensitivity tests}\label{app:gaussian1dextra}

\begin{figure*}
    \centering
    \includegraphics[width=0.7\linewidth]{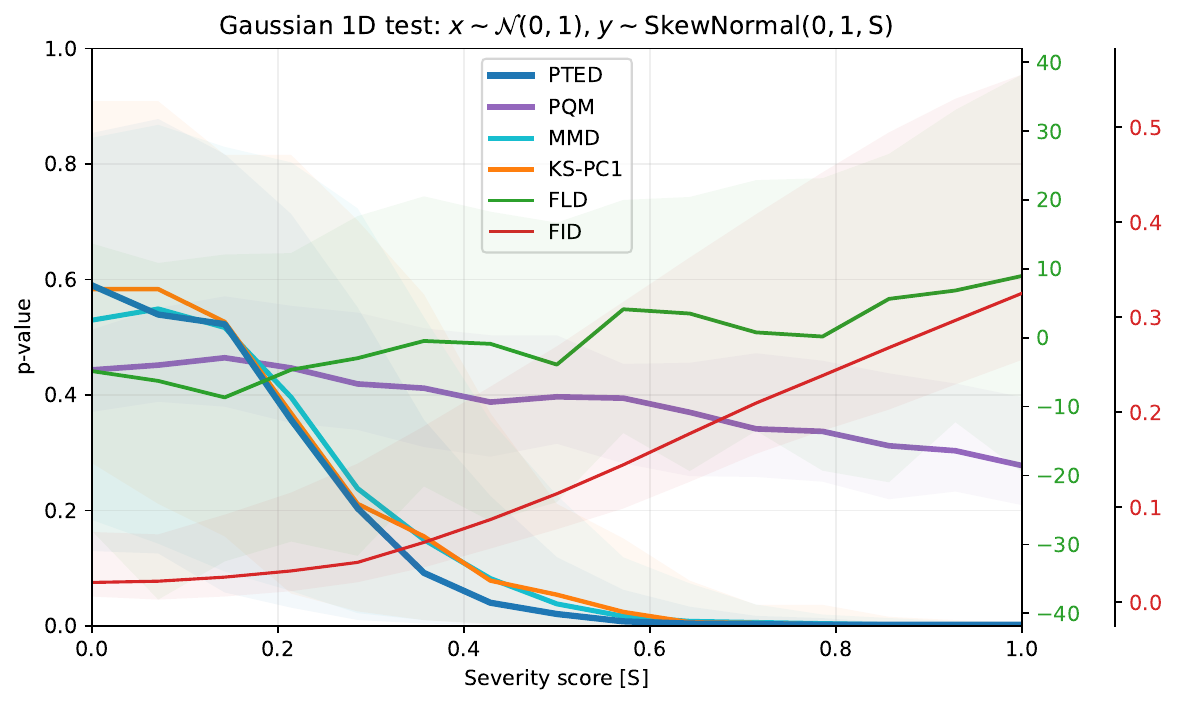}
    \includegraphics[width=0.7\linewidth]{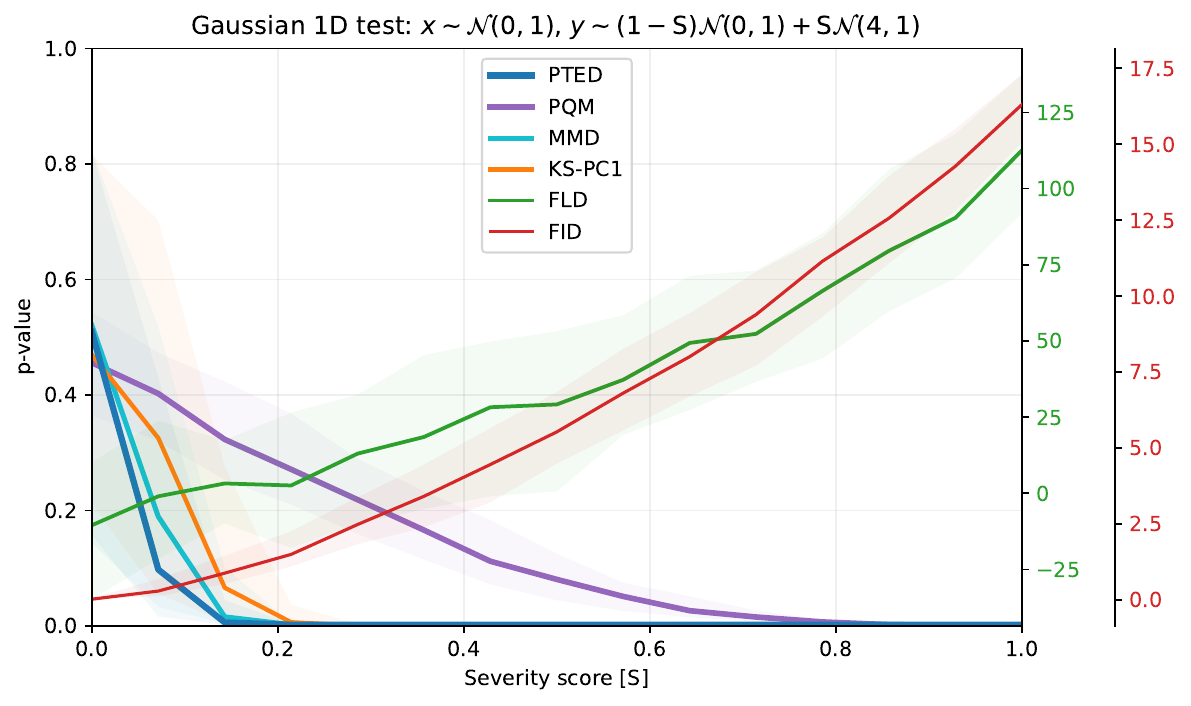}
    \includegraphics[width=0.7\linewidth]{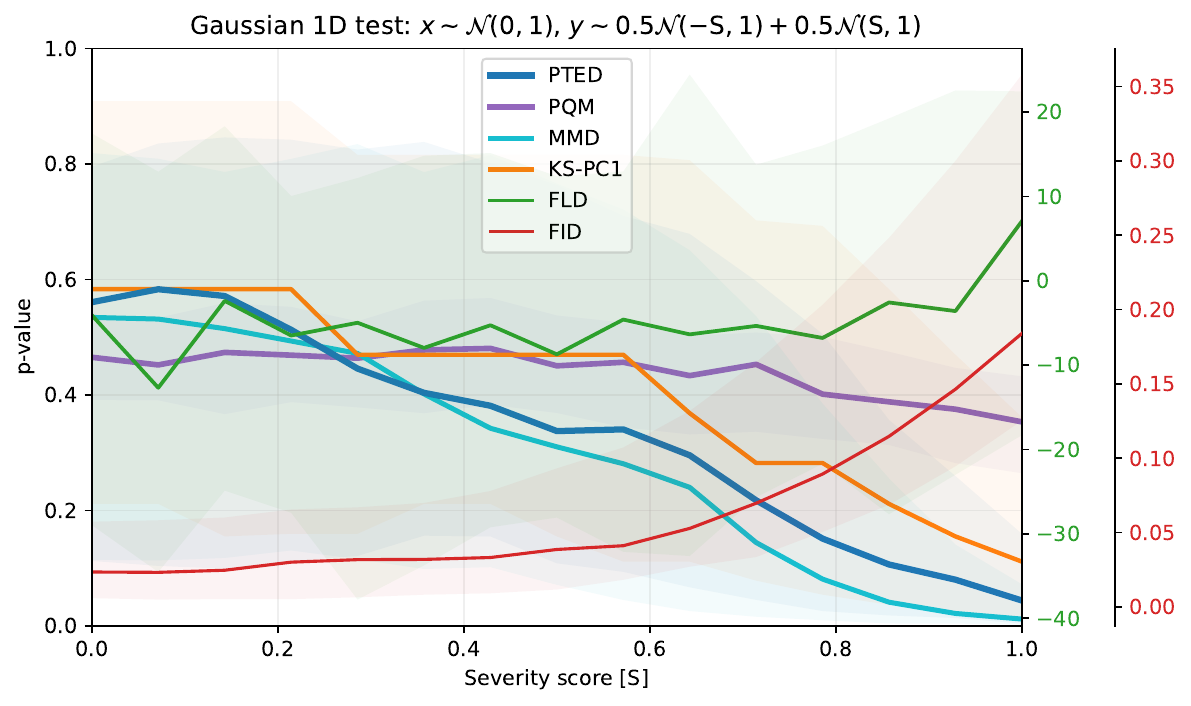}
    \caption{Same as \Fig{gaussian1d} except showing three more tests. Top: the $y$ samples are drawn from a skew-norm distribution with skew equal to $S$. Middle: a second mode with mean 4 is sampled with proportion $S$. Bottom: the $y$ samples are drawn from a bimodal mixture distribution of Gaussians with means $\pm S$. }
    \label{fig:gaussian1dextra}
\end{figure*}

The tests in \Sec{gaussian1d} were chosen to present the clearest picture in the performance comparisons for the various methods.
\Fig{gaussian1dextra} gives a number of other tests which also check for key deviation modes from a pure Normal distribution.
Skewness is the next moment after the mean and variance shifts in \Fig{gaussian1d}.
A contamination test places a second mode at $4\sigma$ representing a failure case where an occasional catastrophic variation occurs (e.g., a hallucination).
The bimodal test checks for the sensitivity to two distributions masquerading as a single more well defined peak, as can be seen in \App{gaussian1d} this issue can be quite subtle.

\section{1D Gaussian sensitivity test visualizations}\label{app:gaussian1d}

\begin{figure*}
    \centering
    \includegraphics[width=0.49\linewidth]{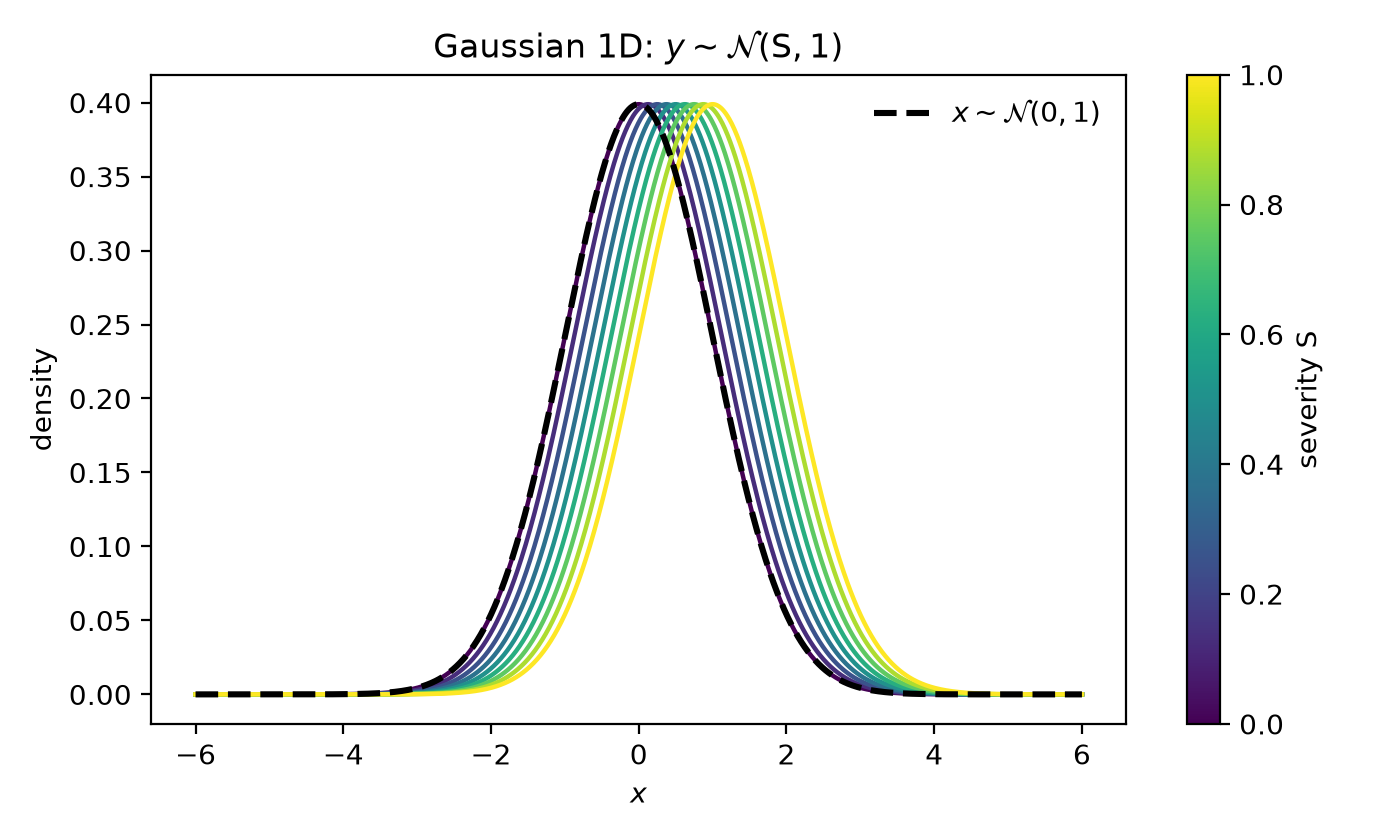}
    \includegraphics[width=0.49\linewidth]{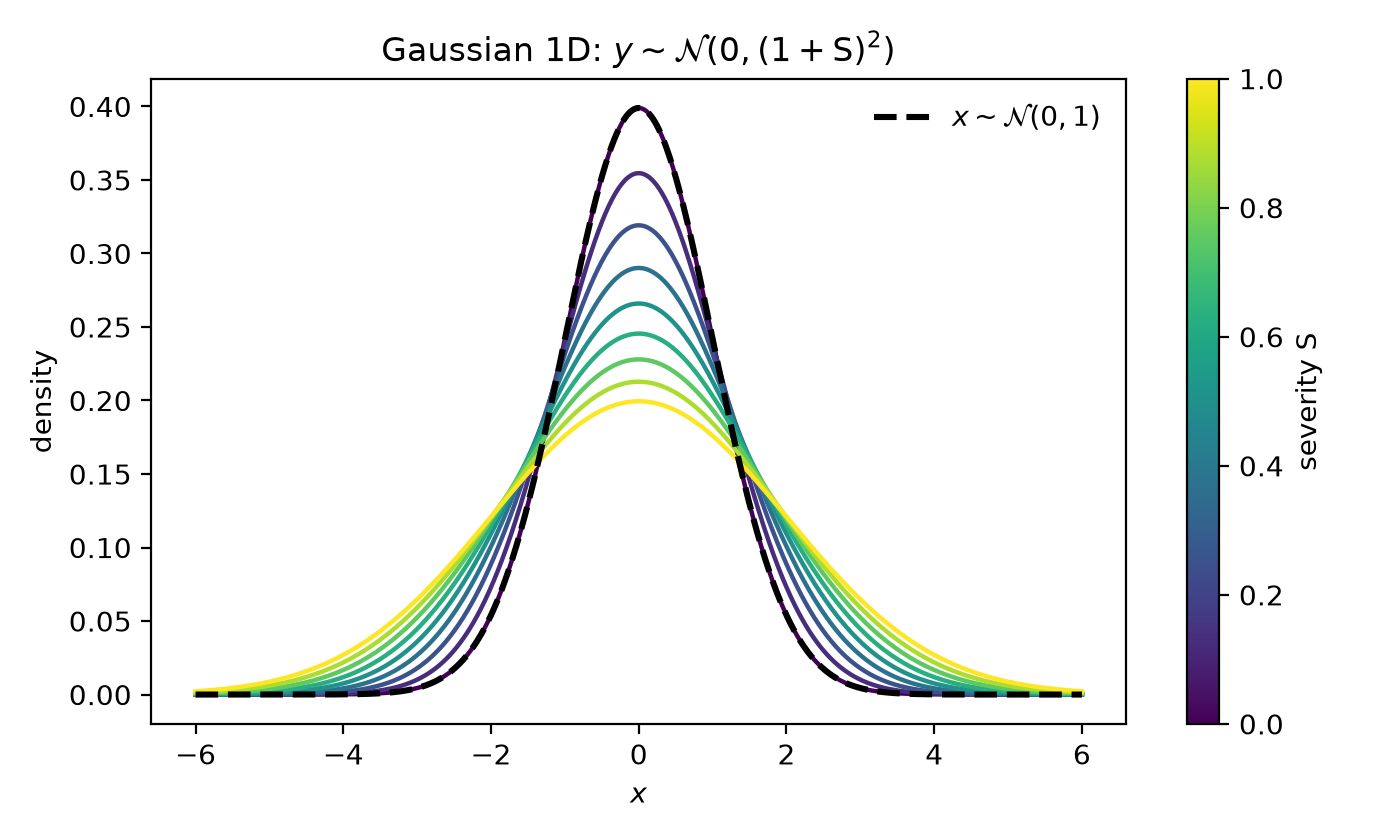}
    \includegraphics[width=0.49\linewidth]{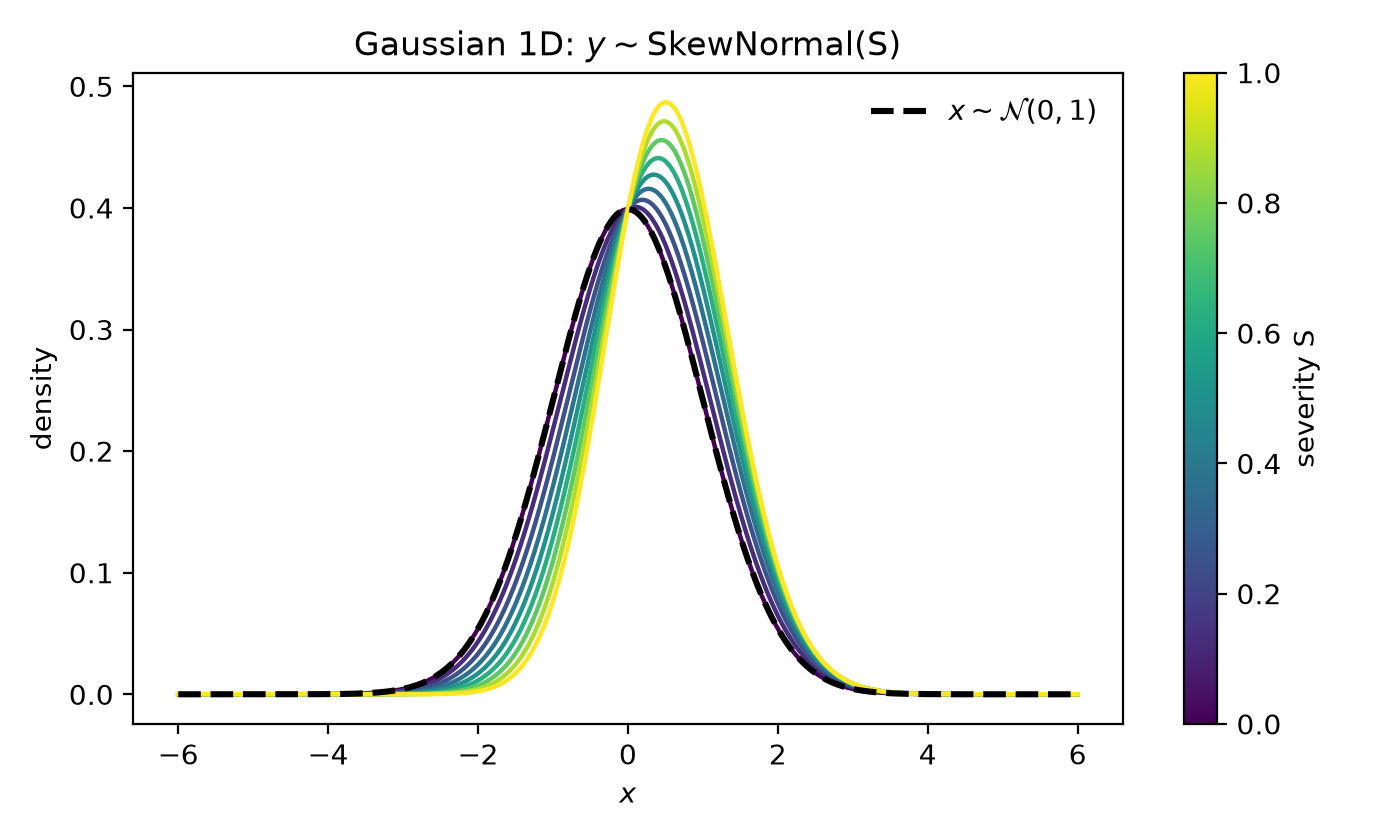}
    \includegraphics[width=0.49\linewidth]{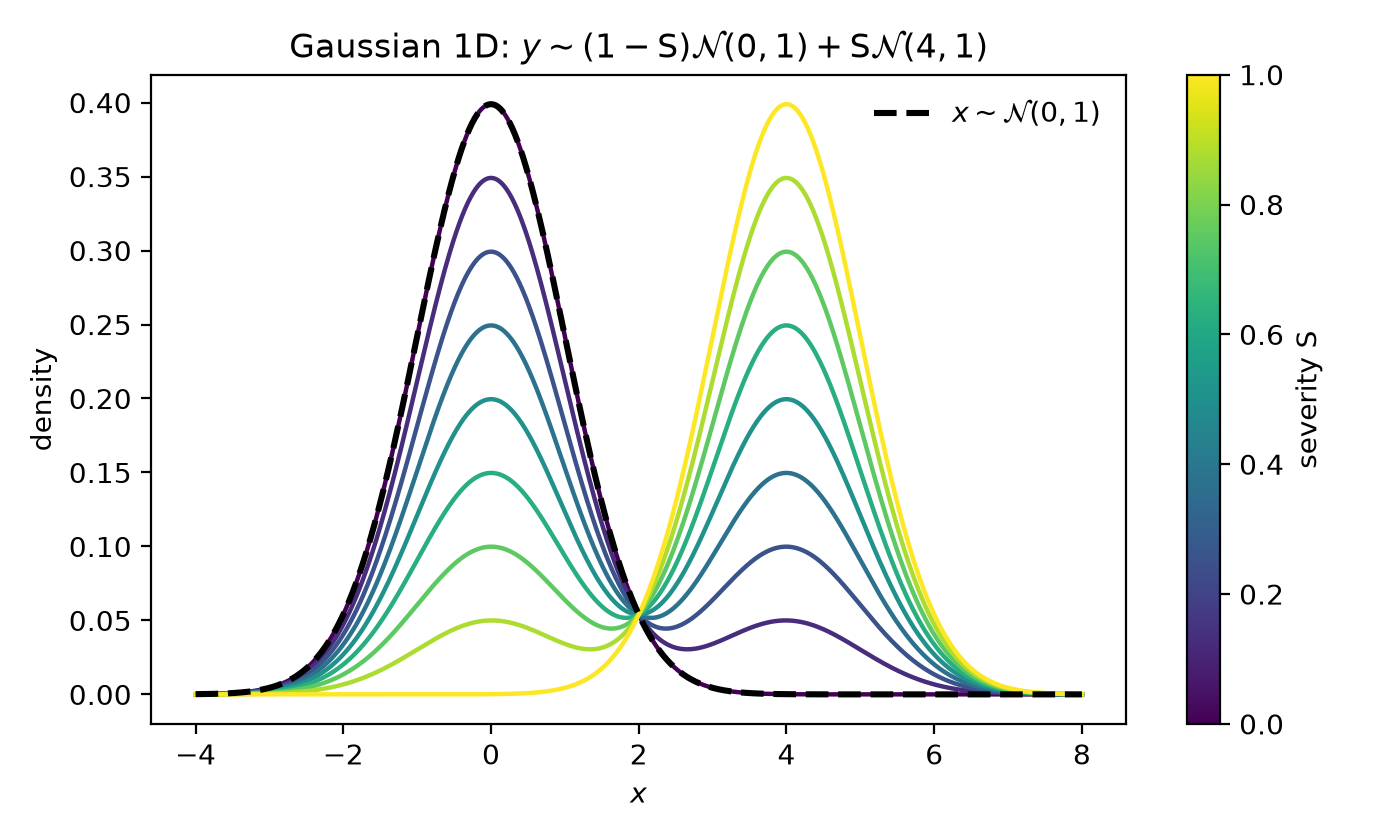}
    \includegraphics[width=0.49\linewidth]{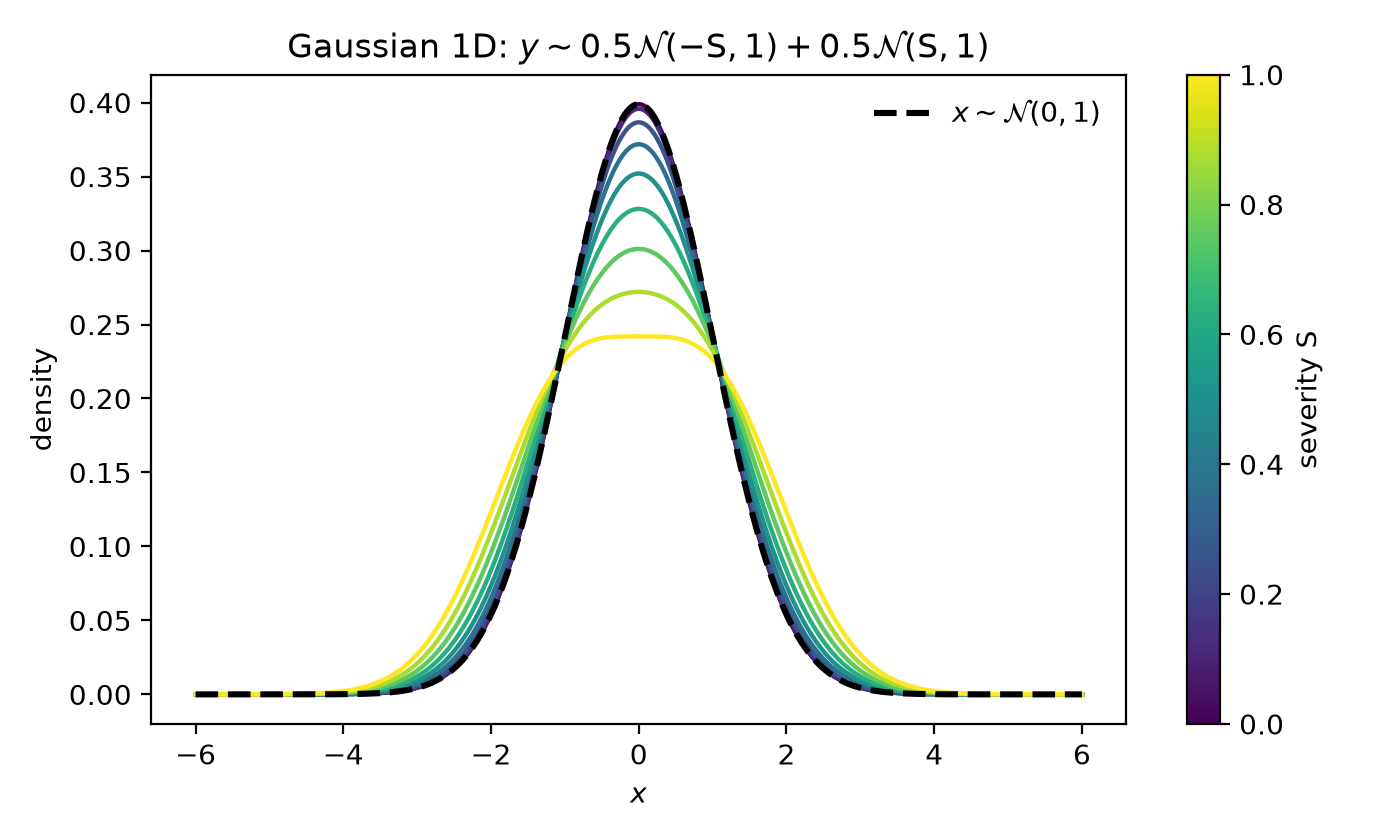}
    \caption{Visualizations of the Gaussian 1D tests in \Fig{gaussian1d} and \Fig{gaussian1dextra}. The sub-figures show the sampling distributions for $x$ and $y$ in each test. The black dashed line gives the $x$ samples distribution which is always a standard Normal distribution. The coloured lines give a progression of severity in the distribution for the $y$ samples. Note that these are the distributions while the tests are always run on 100 samples from these distributions.}
    \label{fig:gaussian1dvis}
\end{figure*}

In \Sec{gaussian1d} a number of deviations from Gaussianity are considered as challenges for multiple two-sample test algorithms.
While the deviations are described mathematically in each title, here each deviation is visualized.
In \Fig{gaussian1dvis} the various deviations are visualized by the density functions that are used to sample the $x,y$ inputs to the two-sample tests.
Coloured by severity it is clear how, with only 100 samples, the distributions would progress from nearly indistinguishable to clearly separate.

\section{MNIST sensitivity tests visualizations}\label{app:mnist}

\begin{figure*}
    \centering
    \includegraphics[width=0.9\linewidth]{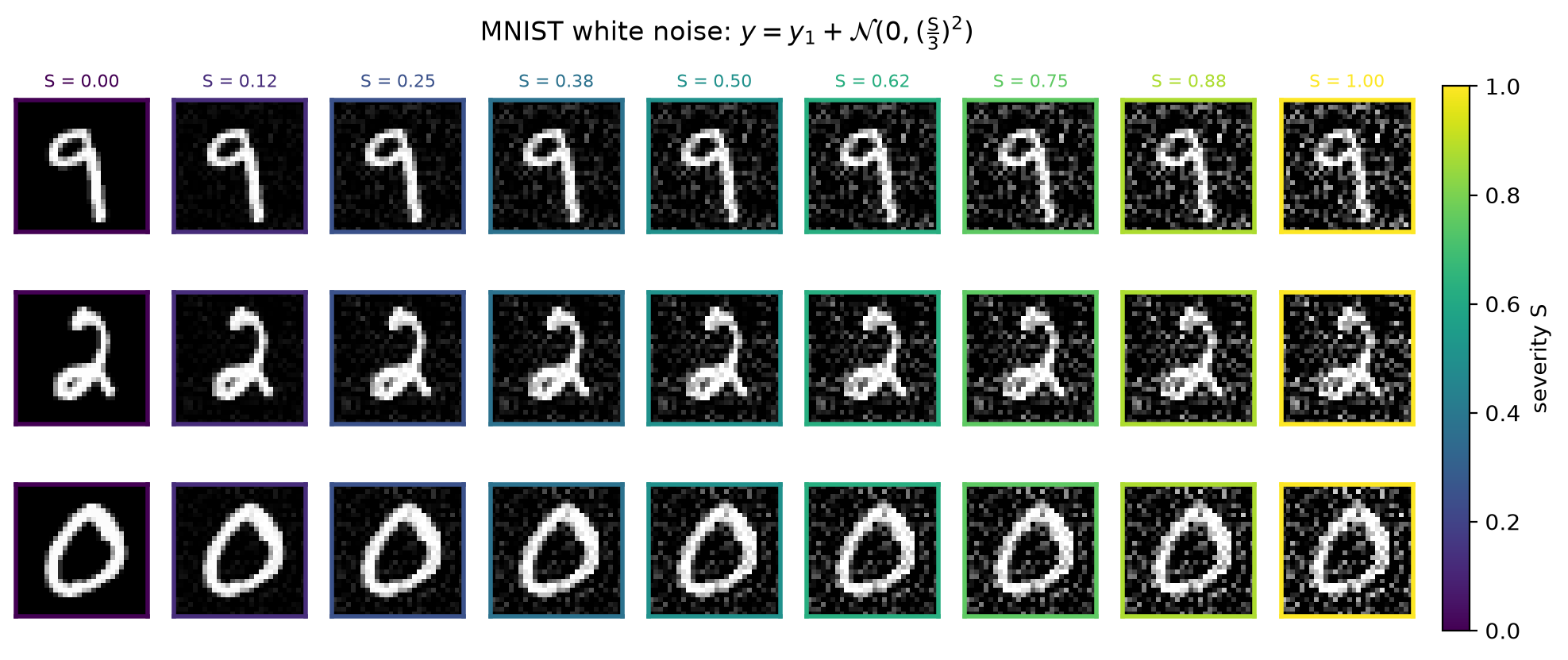}
    \includegraphics[width=0.9\linewidth]{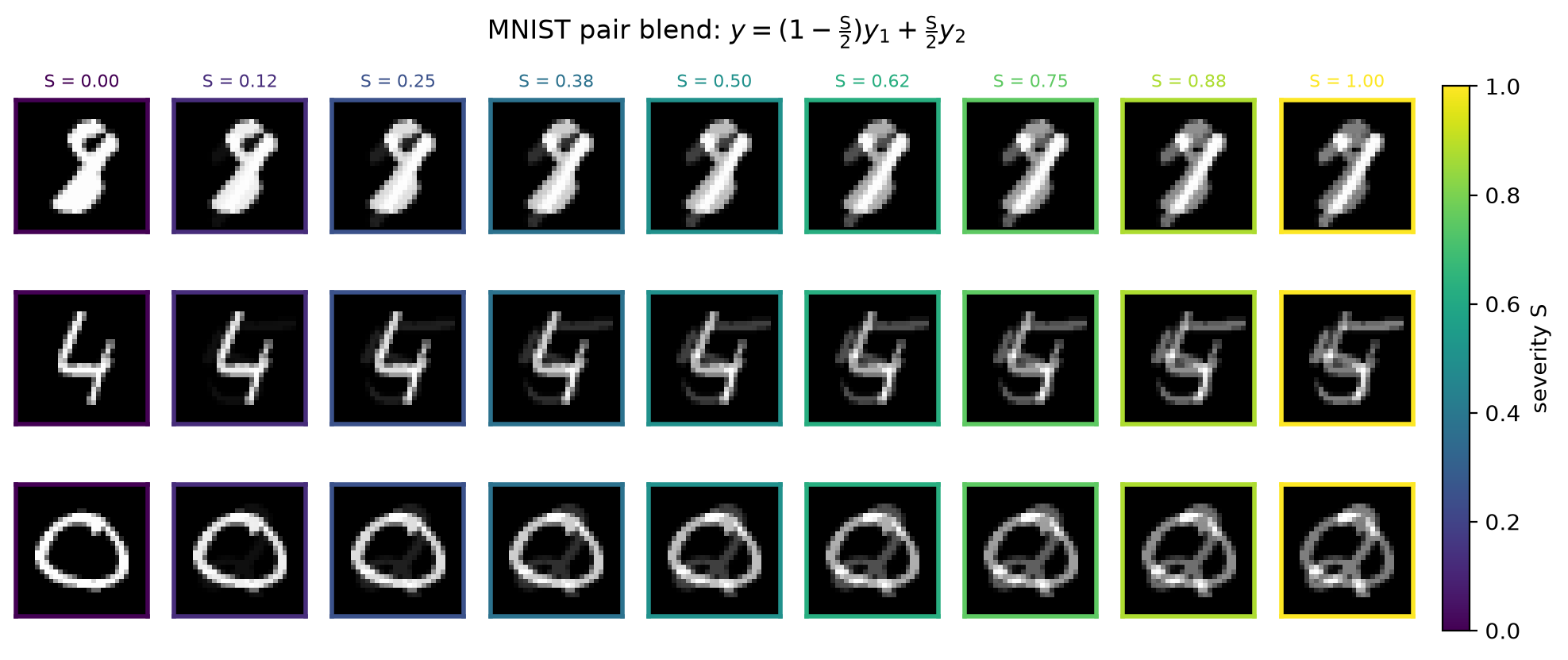}
    \caption{Visualization for two of the MNIST tests in \Fig{mnist} from \Sec{mnist}. The upper figure shows the progression of white noise added to the MNIST digits. The lower figure shows some example blending of two MNIST digits. The ``class drop'' test is not visualized as it was not a modification to the samples, but rather a modification to the probability of drawing a sample from class zero.}
    \label{fig:mnistvis}
\end{figure*}

In \Sec{mnist} the two-sample test algorithms are tested in a high-dimensional setting using the MNIST data.
In \Fig{mnistvis} a visualization is provided to demonstrate how the samples are perturbed before being passed to the various algorithms.
The perturbations are specified in \Sec{mnist}; seeing them makes the sensitivity curves easier to interpret.
For example, in the white noise perturbation examples, it is clear by $S=0.25$ that noise is being added to the images, and sure enough this is when the $p$-value for \PTED drops significantly in \Fig{mnist}.

\section{CIFAR-10 sensitivity tests}\label{app:cifar10}

\begin{figure*}
    \centering
    \includegraphics[width=0.49\linewidth]{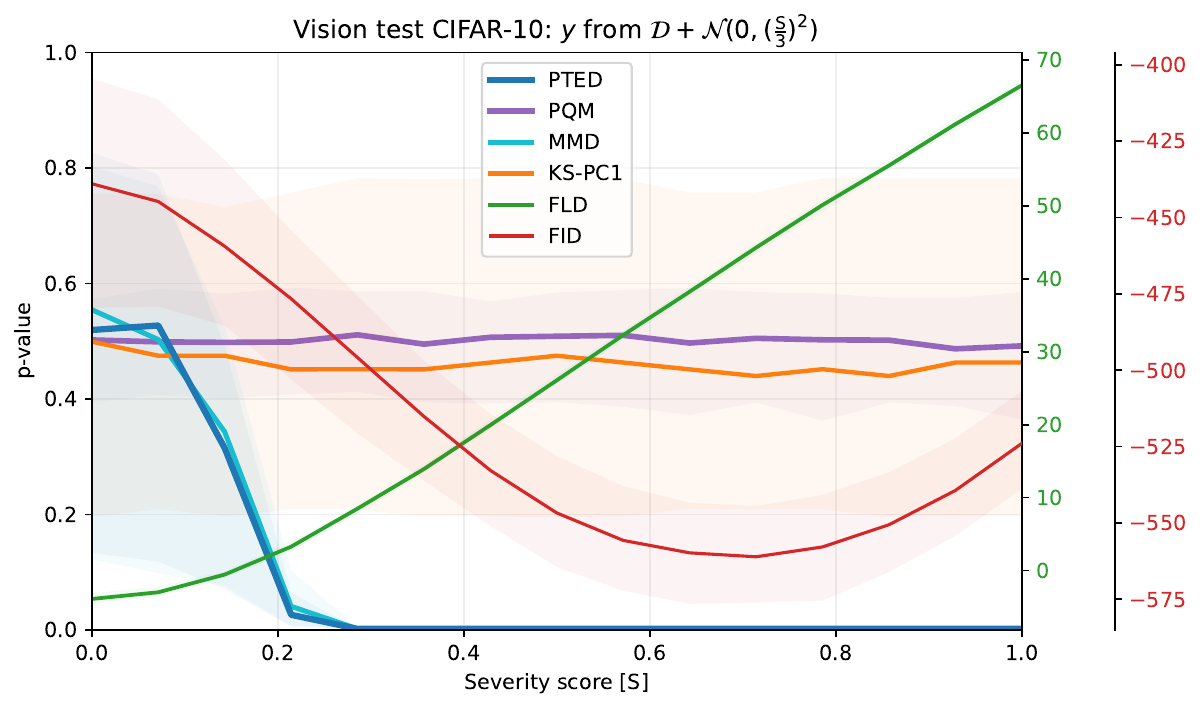}
    \includegraphics[width=0.49\linewidth]{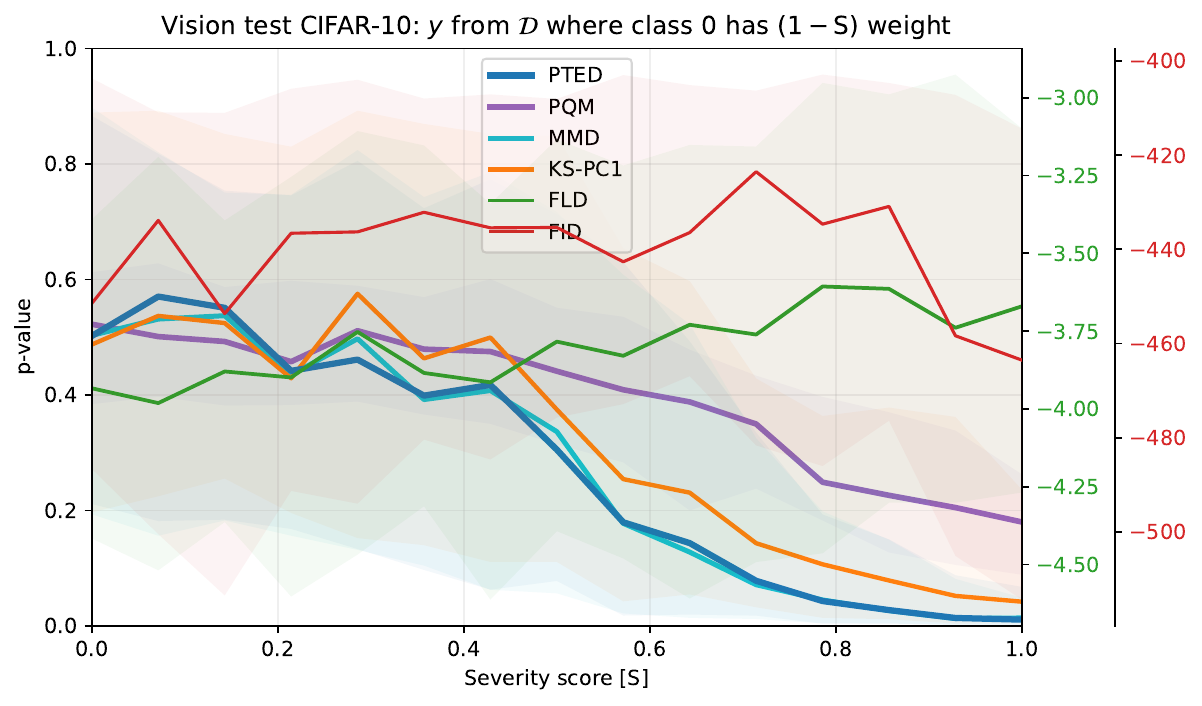}
    \includegraphics[width=0.5\linewidth]{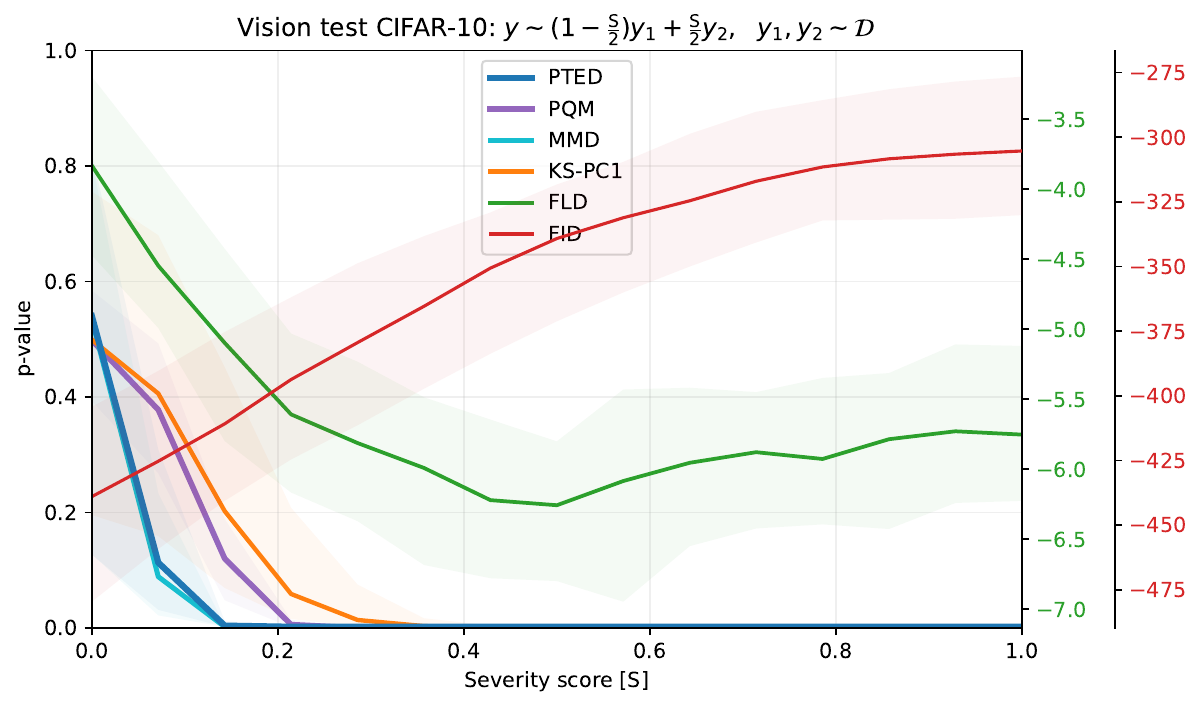}
    \caption{Same as \Fig{mnist} except applied to the CIFAR-10 dataset.}
    \label{fig:cifar10}
\end{figure*}

The CIFAR-10 dataset is a classic machine learning test dataset~\citep{krizhevsky2009learning}, acting as a next step in difficulty after MNIST.
Running the same tests as in \Sec{mnist}, \Fig{cifar10} shows the results are largely similar.
The class-drop perturbation is harder to detect in CIFAR-10 than in MNIST, plausibly because the classes are less separated in the sample space.
The FLD metric again produces lower scores (supposedly better matching) for more severely mismatched samples in the pair blend test.
The sensitivity of FLD to white noise is much less intense for the CIFAR-10 data, likely because there are no single dimensions which have zeros for all examples in a batch (2048).

\section{Probability Integral Transform plots}\label{app:pit}

The Probability Integral Transform (PIT) plot presents a CDF of $p$-values from a multi-test comparison such as a posterior coverage test.
For multiple $p$-values $p_1,\dots,p_{n_{\rm sim}}$ the plot shows the empirical distribution function
\begin{equation}
    \hat{F}(t) = \frac{1}{n_{\rm sim}}\sum_{i=1}^{n_{\rm sim}} \mathbf{1}[p_i \le t] ,
    \label{equ:pit}
\end{equation}
\noindent against the $U(0,1)$ reference $F(t) = t$. 
$\hat{F}$ rising above the diagonal indicates an excess of small $p$-values (over-confidence) and falling below it an excess of large $p$-values (under-confidence).

\begin{figure}
    \centering
    \includegraphics[width=0.32\linewidth]{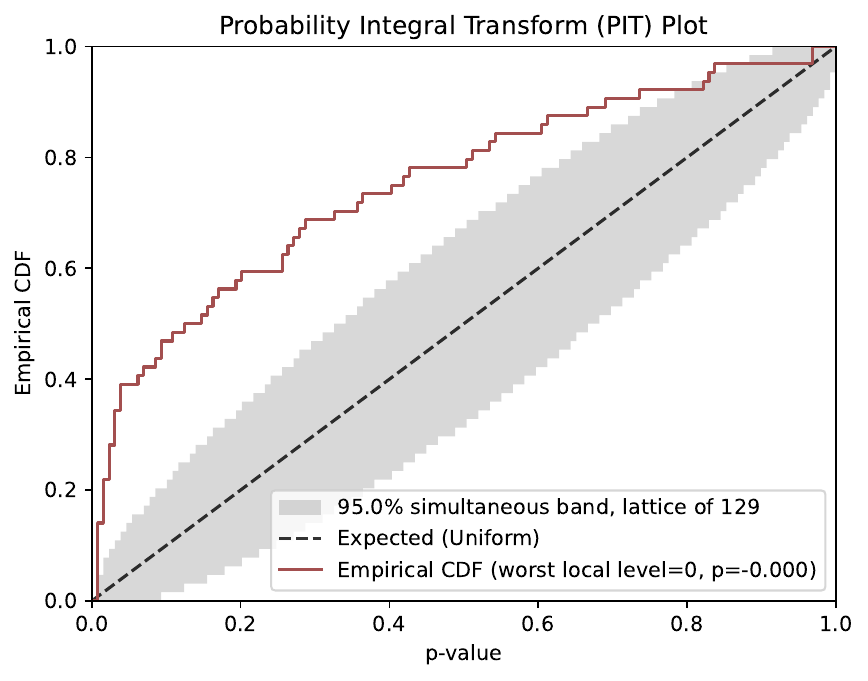}
    \includegraphics[width=0.32\linewidth]{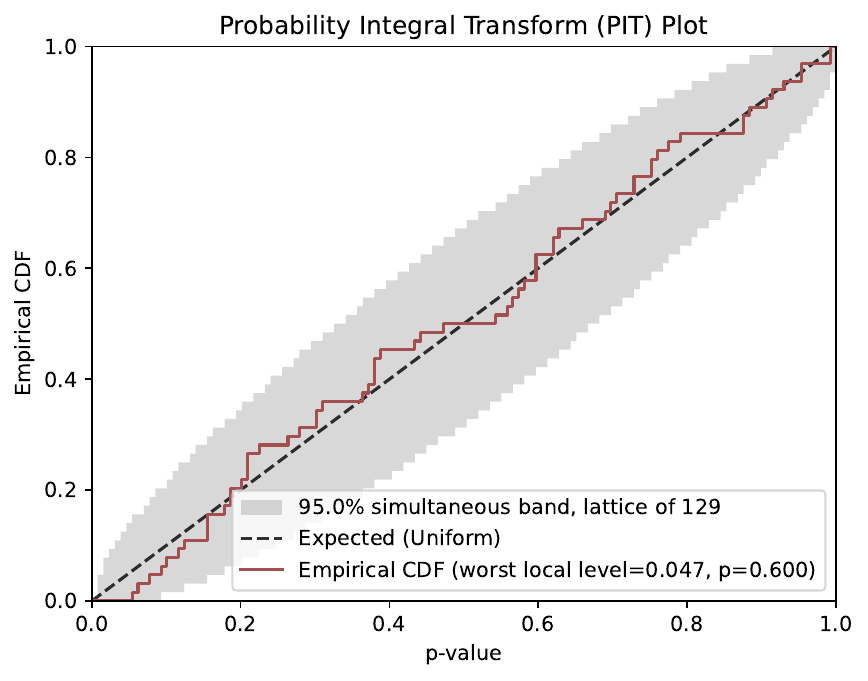}
    \includegraphics[width=0.32\linewidth]{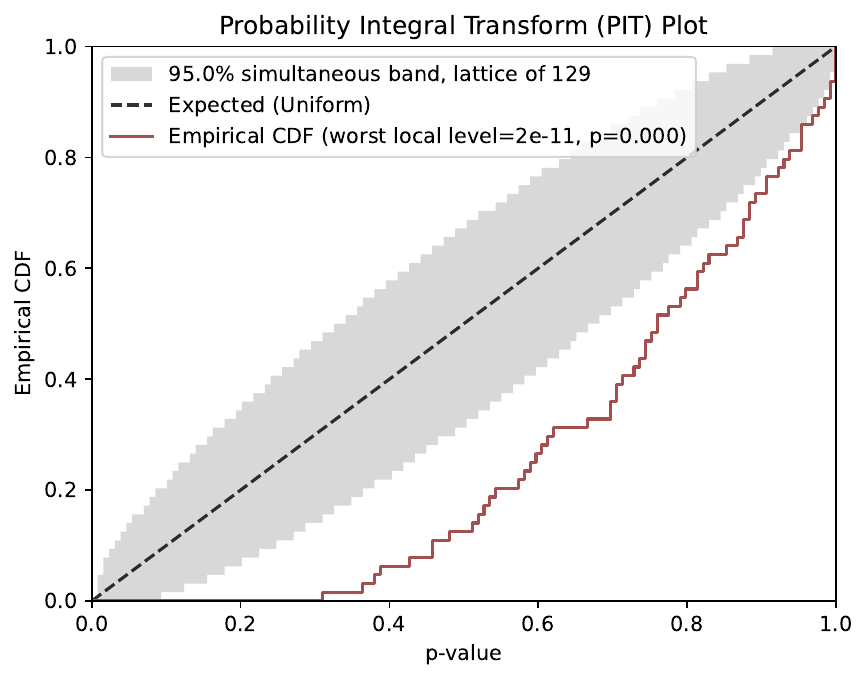}
    \caption{Example \PTED PIT plots for \Sec{coverage} showing overconfident (left), well calibrated (centre), and under-confident (right) posterior results. The shaded region shows the 95\% containment band under the null, any part of the curve extending beyond the grey region indicates a null rejection at 95\% significance.}
    \label{fig:pit}
\end{figure}

This is a useful visualization for spotting pathological behaviours that may not get picked up by a summary value like that from the Fisher $p$-value combination.
In \Fig{pit}, three example PIT plots are shown for overconfident, well calibrated, and under-confident posteriors.
In the overconfident case, there is an excess of low $p$-values causing the curve to rise significantly above the 1:1 line.
In the well calibrated case, the curve follows the 1:1 line closely.
The shaded region is a simultaneous band, so a curve that stays inside it is consistent with $H_0$ at the stated level.
In the under-confident case, there is an excess of high $p$-values causing the curve to fall significantly below the 1:1 line.

The PIT plot shows potential features that an aggregate measure can miss.
Excessive structure in the curve may indicate there are regions of parameter space which are poorly calibrated even if the average performance is good.
Read together, the Fisher aggregate, the PIT curve and a direct visualization of the samples give a reasonably complete picture; no combination of them, however, can establish $H_0$, only fail to reject it.

\end{document}